\documentclass[aps,pre,twocolumn,nofootinbib,tightenlines,superscriptaddress,nopacs,amsmath,amssymb,final,letterpaper]{revtex4-1}

\usepackage[utf8]{inputenc}
\usepackage{calc}
\usepackage{graphicx}
\usepackage{amsmath,amssymb,amsthm}
\usepackage{bbold}
\usepackage{bm}
\usepackage{color}
\usepackage[dvipsnames]{xcolor}
\usepackage{enumitem}
\usepackage{tikz}
\usepackage{float}

\newcommand{\abs}[1]{\vert #1\vert}
\newcommand{\overbar}[1]{\mkern 1.5mu\overline{\mkern-1.5mu#1\mkern-1.5mu}\mkern 1.5mu}

\def\multiset#1#2{\ensuremath{\left(\kern-.3em\left(\genfrac{}{}{0pt}{}{#1}{#2}\right)\kern-.3em\right)}}

\usepackage[colorlinks=true,linkcolor=blue,urlcolor=blue,citecolor=blue,anchorcolor=blue]{hyperref}

\renewcommand{\thefigure}{\textbf{\arabic{figure}}}
\renewcommand{\thetable}{\textbf{\arabic{table}}}

\begin{document}
\title{Network mutual information measures for graph similarity}

\author{Helcio \surname{Felippe}}
\affiliation{Department of Network and Data Science, Central European University, Vienna, Austria}

\author{Federico \surname{Battiston}}
\email{battistonf@ceu.edu}
\affiliation{Department of Network and Data Science, Central European University, Vienna, Austria}

\author{Alec \surname{Kirkley}}
\email{alec.w.kirkley@gmail.com}
\affiliation{Institute of Data Science, University of Hong Kong, Hong Kong SAR, China}
\affiliation{Department of Urban Planning and Design, University of Hong Kong, Hong Kong SAR, China}
\affiliation{Urban Systems Institute, University of Hong Kong, Hong Kong SAR, China}

\date{\today}


\begin{abstract}
    A wide range of tasks in network analysis, such as clustering
    network populations or identifying anomalies in temporal graph
    streams, require a measure of the similarity between two graphs.
    To provide a meaningful data summary for downstream scientific
    analyses, the graph similarity measures used for these tasks must
    be principled, interpretable, and capable of distinguishing
    meaningful overlapping network structure from statistical noise at
    different scales of interest.  Here we derive a family of graph
    mutual information measures that satisfy these criteria and are
    constructed using only fundamental information theoretic
    principles. Our measures capture the information shared among
    networks according to different encodings of their structural
    information, with our mesoscale mutual information measure
    allowing for network comparison under any specified network
    coarse-graining. We test our measures in a range of applications
    on real and synthetic network data, finding that they effectively
    highlight intuitive aspects of network similarity across scales in
    a variety of systems.   
\end{abstract}

\maketitle

\section{Introduction}

Network similarity and distance measures are widely applied across
science and engineering disciplines for understanding the shared
structure among multiple
graphs~\cite{sharan2006modeling,nikolova2003approaches,wang2020measurement,zeng2009comparing,guo2021semantic,kriege2020survey}.
Common graph-level analysis tasks such as network population
clustering~\cite{de2015structural}, network
regression~\cite{ok2020graph}, and network
classification~\cite{attar2017classification} require as input a
network similarity measure and are highly sensitive to this choice,
leading to the construction of a vast number of graph similarity
measures to accommodate different application
needs~\cite{vishwanathan2010graph,wills2020metrics,hartle2020network}.
blackIn the exploratory data analysis setting, the goal of performing
graph similarity or distance calculations is often to identify some
meaningful summary of a set of graphs or network layers---for example,
clusters of graphs or a subset of anomalous
graphs~\cite{kyosev2020measuring,ranshous2015anomaly}. In this case,
it is essential that the graph similarity or distance measure used is
principled, interpretable, and capable of distinguishing meaningful
shared structure among the input graphs from statistical noise.
Existing measures based on collections of network summary statistics
or feature embeddings~\cite{roy2014modeling,soundarajan2014guide} are
often hard to interpret, and there is no clear criterion for which
features to include in the similarity/distance calculation. Methods
based on graph spectra~\cite{apolloni2011introduction,wilson2008study}
have a clear connection to community structure and random walk
dynamics on graphs, but are also challenging to interpret given the
localization tradeoff between graph space and frequency
space~\cite{hammond2011wavelets}. Additionally, many of these methods
that do not rely on external labels do not explicitly aim to capture
statistically meaningful structural distinctions between graphs,
meaning that they can reflect a high amount of similarity between two
graphs that is purely due to constraints imposed by global properties
such as graph density or degree distributions. In this context,
information theory can provide a principled framework for deciding the
extent to which two networks are similar by allowing us to compute a
mutual information between two graphs that quantifies the amount of
information they share under a particular encoding scheme. Mutual
information measures resulting from a minimum description length (MDL)
approach have been used to compare partitions of objects in a diverse
range of
applications~\cite{newman2020improved,mcdaid2011normalized,vinh2010information,kirkley2022spatial,kirkley2024dynamic},
becoming the standard measure for comparing partitions of networks in
the community detection setting~\cite{fortunato2016community}. 

\begin{figure*}[ht]
\includegraphics[scale=0.90]{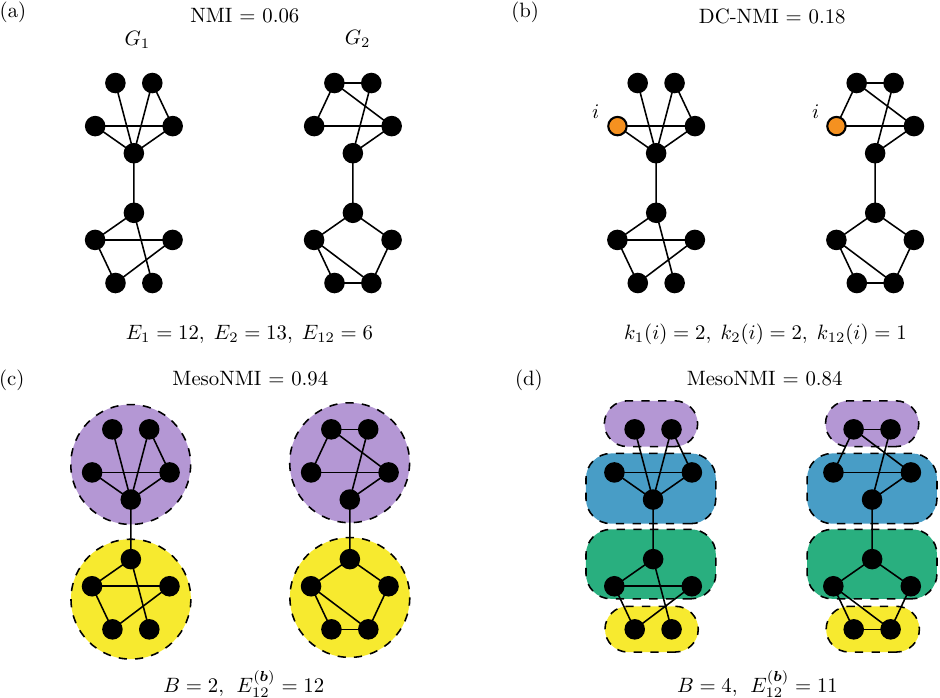}
\caption{
    \textbf{
    Family of proposed network mutual information measures for graph similarity. 
    }
    (a)~Standard normalized mutual information (NMI,
    Eq.~\ref{eq:NMIgraph}) between networks $G_1$ and $G_2$, with
    shared node labels indicated by node positions. Due to little
    overlap in the edge positions, this NMI measure returns a score
    $\text{NMI}(G_1;G_2)\approx 0$. 
    (b)~Degree-corrected normalized mutual information (DC-NMI,
    Eq.~\ref{eq:DCNMI}) between graphs $G_1$ and $G_2$. Due to little
    overlap in node neighborhoods, we also see a low value for
    $\text{DC-NMI}(G_1;G_2)$. 
    (c)~Mesoscale normalized mutual information (MesoNMI,
    Eq.~\ref{eq:mesoNMI2}) between networks $G_1$ and $G_2$ with
    respect to the indicated node partition $\bm{b}$ into $B=2$ groups
    (colored, circled sets of nodes). Since the mesoscale structure of
    these networks is quite similar, as indicated by the edge overlap
    $E^{(\bm{b})}_{12}=12$, we have a high similarity value
    $\text{MesoNMI}^{(\bm{b})}(G_1;G_2)\approx 1$. 
    (d)~MesoNMI between the two networks but with respect to a
    different partition $\bm{b}$ with $B=4$ groups (colored, circled
    sets of nodes). Here we still see a relatively high MesoNMI value,
    indicating substantial shared structure at this smaller scale. For
    reference, the Jaccard index among the edge sets in this case is
    $\abs{G_1\cap G_2}/\abs{G_1\cup G_2}=0.315$---a much higher value
    than the NMI in panel (a)---indicating that the edge overlap is
    not much different than expected based on the network densities.
}
\label{fig:fig1_diagram}
\end{figure*}

An often overlooked aspect of graph similarity measures is what scale
is highlighted in the network comparison calculation. Most existing
graph similarity measures can be broadly categorized based on the
structural level---micro, meso, or macro---at which they highlight
network similarity~\cite{soundarajan2014guide}. For unsupervised tasks
involving graph similarity measures, any of the three network scales
may be of interest. For example, when comparing snapshots of a
longitudinal social network to examine its temporal evolution,
different analyses may require a network similarity measure that
focuses on micro-scale structural overlap among individuals' ego
networks, a measure that captures the evolving meso-scale community
structure of the network, or a measure that tracks macro-scale summary
statistics of the network such as its total density. Existing graph
similarity measures have largely been constructed to highlight a
particular scale of interest---for example, individual edge overlap at
the microscale is the focus of traditional graph edit
distances~\cite{gao2010survey}, and global path structure in the
network is highlighted by betweenness or closeness centrality-based
network similarity measures~\cite{roy2014modeling}. Spectral distance
measures can simultaneously highlight different scales within the
network due to the presence of high and low frequency
eigenmodes~\cite{apolloni2011introduction,wilson2008study}, but it is
unclear exactly to what extent each of the three scales contributes to
the computed similarity. There is currently a lack of graph similarity
measures that have a clear dependence on the scale of interest, which
is critical for interpretability and flexibility in applications.

In~\cite{coupette2021graph}, an information theoretic measure for
comparing networks is presented that is based on identifying shared
substructures (e.g.~stars and cliques) that minimize the description
length of a pair of networks. This approach is quite elegant and
flexible but is limited in practice to subgraph similarities and can
become computationally burdensome with the inclusion of larger
structural vocabularies. It also does not allow for the comparison of
graphs at the meso- or macro-scale while ignoring edge-level details,
as is possible with the mesoscale mutual information measure presented
in this paper. A pairwise graph mutual information measure is proposed
in~\cite{ESCOLANO201712}, but this measure requires a continuous
embedding of the input graphs which may produce distortions of its
topological structure. It also does not consider the problem of
adjusting the scale at which the graph structure is compared.
In~\cite{kirkley2023compressing}, an MDL approach for clustering
populations of node-aligned networks is presented, which is equivalent
to maximum a posteriori Bayesian inference with the model of
\cite{YKN22} up to a choice of prior.  This approach considers
encoding sample graphs based on their cluster's representative
``mode'' network. The encoding for a cluster of networks resembles the
encoding used here for the standard (non-degree-corrected) conditional
entropy between graphs, but it does not accommodate the comparison of
networks in a symmetric pairwise manner, and also does not address the
issue of scale.  In general, principled approaches to measure graph
similarity in the node-aligned setting that we consider here have
primarily focused on analyzing entire populations of
graphs~\cite{lunagomez2021modeling,coupette2022differentially},
leaving a relative scarcity of methods for principled pairwise network
comparison.

In this paper we construct a family of mutual information measures for
computing graph similarity at different scales (see
Fig.~\ref{fig:fig1_diagram}). We do this by considering multiple
encodings of node-aligned graphs that exploit different aspects of
shared network structure---edges, node neighborhoods, and mixing
patterns among arbitrary partitions of nodes in a network---and using
these encodings to quantify the amount of shared information between
two input graphs.  Our measures are principled, interpretable, fast to
compute, and naturally highlight the significant shared structure
between two graphs beyond the overlap expected due to global
structural constraints such as edge density and degree distributions.
To validate our methods we apply them in a range of experiments,
finding that they intuitively capture structural perturbations in
synthetic networks and identify meaningful shared structure across
layers in multilayer networks.


\section{Methods}

To measure the similarity between two graphs $G_1$ and $G_2$, we can
appeal to information theory and compute the amount of information
shared between $G_1$ and $G_2$, also known as their mutual
information~\cite{cover2012elements}. Due to nice properties such as
non-negativity, boundedness, and symmetry, mutual information measures
and their normalized variants have been widely used across
unsupervised learning tasks within and outside of network
science~\cite{vinh2010information,lancichinetti2008benchmark,kirkley2022spatial}. 

A key component in the development of a mutual information measure is
the specification of an \emph{encoding scheme}, which specifies how
one will represent data configurations in a codebook that is used to
communicate the data to a receiver. Encoding schemes that focus on
different structural properties in a graph will naturally tend to
result in different estimates of the information shared between the
graphs, thus different values of the mutual information. Here we
describe three encoding schemes that aim to highlight shared structure
among graphs at different scales, and discuss how to use them to form
three mutual information measures that capture different aspects of
similarity among networks. 

\subsection{Network mutual information}
\label{sec:MIdyadic}

Let $G_1$ and $G_2$ be graphs on the same set of $N$ labelled nodes.
We will focus our attention on the case where these graphs are
undirected with no self- or multi-edges, but discuss later on how our
measures can be straightforwardly extended when we relax these
assumptions. For convenience we will represent $G_1$ and $G_2$ with
sets of undirected, unweighted edges of sizes $E_1$ and $E_2$ (the
number of unique edges in $G_1$ and $G_2$) respectively. In principle,
our measure can handle any two graphs $G_1$ and $G_2$ that have the
same number of nodes $N$---regardless of whether these node labels are
aligned. However, in the case of unaligned node labels one must
perform graph alignment prior to using the algorithm in order to
obtain meaningful results, and this is a highly nontrivial task that
will have a substantial impact on the similarity value.  Therefore,
our measures will primarily be of interest for comparing node-aligned
networks generated from cross-sectional studies with identical
constraints across subjects (e.g. brain networks among a set of
patients~\cite{sporns2010networks}), longitudinal studies (e.g.
multiple observations of a social network among the same set of
students~\cite{eagle2006reality}), and experimental studies with
repeated measurements~\cite{newman2018network}. We will let
$\mathcal{G}$ be the set of ${N\choose 2}$ possible edges on these $N$
nodes, so that $G_1,G_2\subseteq \mathcal{G}$. 

Analogous to constructing a mutual information among
labellings~\cite{newman2020improved}, we will derive our network
mutual information measure by first considering the information
required to transmit the network $G_2$ to a receiver using a
particular encoding scheme. Then we will see by how much we can
decrease our information cost if we transmit $G_1$ and exploit the
graphs' shared structure to transmit $G_2$. The information savings we
incur is the mutual information between $G_1$ and $G_2$ under the
specified encoding. 

We will assume that the receiver knows the number of nodes $N$ and
edges $E_2=\abs{G_2}$ in the second graph. (Transmitting this quantity
will require comparatively negligible information, so we can ignore it
anyway.) There are ${N \choose 2}$ possible unique undirected edges in
$\mathcal{G}$ that one can construct using $N$ nodes, so there are
${{N \choose 2}\choose E_2}$ possible networks $G_2$ consistent with
the constraints known by the receiver, among which we must specify a
single one to transmit the graph structure $G_2$. The base-two
logarithm of this quantity is therefore the approximate number of bits
we will need to encode all these possibilities using binary strings if
we choose a simple encoding that does not involve transmitting any
intermediate summary information about $G_2$ to the receiver. (We will
describe two possibilities for such an encoding later on.) We call
this quantity of information the \emph{entropy} $\text{H}(G_2)$ of the
graph $G_2$ under this simple encoding scheme, and it is given by 
\begin{align}\label{eq:H2}
    \text{H}(G_2) = \log{{N \choose 2}\choose E_2}. 
\end{align}
For simplicity of presentation, we will assume all logarithms are
base-two for the remainder of the paper. 

We can simplify Eq.~(\ref{eq:H2}) into a more recognizable form using the
Stirling approximation \mbox{$\log x! \approx x\log x - x/\text{ln}(2)$},
giving
\begin{align}\label{eq:H2canonical}
    \text{H}(G_2) \approx {N\choose 2}\text{H}_b(p_2),    
\end{align}
where $p_2=E_2/{N\choose 2}$ is the fraction of all possible edges
occupied in the graph and 
\begin{align}
    \text{H}_b(p) = -p\log p - (1-p)\log(1-p)    
\end{align}
is the binary entropy function. Equation~(\ref{eq:H2canonical}) tells
us that it takes approximately $\text{H}_b(p_2)$ bits of information
to specify the existence or non-existence of an edge for each of the
${N\choose 2}$ possible edge slots, given that the receiver knows
there will be $E_2$ total edges.

Now, consider the case where the receiver already knows $G_1$ prior to
us transmitting $G_2$. In this case, the information required to
transmit $G_2$ should be reduced, since we can exploit the information
shared between $G_1$ and $G_2$ to reduce the size of our encoding.
More specifically, this reduction of information is possible when the
receiver knows both $G_1$ \emph{and} how $G_2$ differs from $G_1$,
since this additional constraint further reduces the number of
possibilities for $G_2$ that need to be encoded. In the standard
formulation of mutual information between labellings, the
discrepancies between the labellings are encoded in a
\emph{contingency table}, which counts the number of instances in
which an object is classified into one cluster under the first
labelling and another cluster in the second labelling. Typically one
ignores the amount of information required to transmit the contingency
table between two labellings to the receiver, as its information
content vanishes in the limit of large labellings~\cite{newman2020improved}.

We can consider the labelling associated with the edge set $G_i$ as a
length-${N\choose 2}$ binary vector whose indices represent all
possible edges in $\mathcal{G}$, and that contains a $1$ for all
entries corresponding to the edges that are present in $G_i$.
(Equivalently, one can just consider flattening the upper triangle of
the adjacency matrix representation of $G_i$ into a vector.) In this
case, the contingency table comparing the labels in $G_1$ and $G_2$
takes a particularly simple form. For any possible edge $(i,j)$ with
$1\leq i,j\leq N$, we can either have that:
\begin{enumerate}
    \item $(i,j) \in G_1$ and $(i,j)\in G_2$. We call this subset of
    edges \emph{true positives} to indicate that they are contained in
    both sets $G_1$ and $G_2$. The set of true positives is given by
    $G_1\cap G_2$, whose size we denote with $E_{12}=\abs{G_1\cap
    G_2}$. 
    \item $(i,j) \in G_1$ and $(i,j)\notin G_2$. We call this subset
    of edges \emph{false negatives} to indicate that they are
    contained in the set $G_1$ but not the set $G_2$. The set of false
    negatives is given by $G_1\setminus G_2$, whose size is
    $E_1-E_{12}$.  
    \item $(i,j) \notin G_1$ and $(i,j)\in G_2$. We call this subset
    of edges \emph{false positives} to indicate that they are not
    contained in the set $G_1$ but are contained in the set $G_2$. The
    set of false positives is given by $G_2\setminus G_1$, whose size
    is $E_2-E_{12}$.  
    \item $(i,j) \notin G_1$ and $(i,j)\notin G_2$. We call this
    subset of edges \emph{true negatives} to indicate that they are
    not contained in either of $G_1$ or $G_2$. The set of true
    negatives is given by $\mathcal{G}-(G_1 \cup G_2)$, whose size is
    ${N \choose 2}-E_1-E_2+E_{12}$.  
\end{enumerate}
These true/false positives/negatives are the four values of the
contingency table between the binary labellings associated with $G_1$
and $G_2$.

Given that the receiver knows $G_1$, $E_2$, and the contingency
table---or, equivalently, the number of true positives $E_{12}$ since
this is the only linearly independent unknown---we can compute the
logarithm of the number of possible configurations that $G_2$ can take
under these constraints as the \emph{conditional entropy}
$\text{H}(G_2\vert G_1)$. This is given by
\begin{align}\label{eq:H21}
    \text{H}(G_2\vert G_1) = \log {E_1 \choose E_{12}}{{N\choose 2}-E_1 \choose E_2 - E_{12}}.    
\end{align}
The first term in the product counts the number of ways to choose the
$E_{12}$ true positives from the set of edges in $G_1$, and the second
term counts the number of ways to choose the $E_2-E_{12}$ false
positives from the remaining edges not contained in $G_1$. The two
actions together, which fully specify $G_2$, can be taken
independently and so the total number of combinations available is
given by the product of the two binomial coefficients.

We can now quantify the amount of information shared between the
graphs $G_1$ and $G_2$ as the reduction in the entropy of $G_2$ that
results from knowing $G_1$ and the contingency table. We call this the
\emph{mutual information} between the graphs, and it can be
represented mathematically as
\begin{align}\label{eq:MIH2-H21}
    \text{MI}(G_1,G_2) &= \text{H}(G_2) - \text{H}(G_2\vert G_1).
\end{align}
Grouping terms and applying Stirling's approximation, we arrive at a
simple, manifestly symmetric form for the graph mutual information: 
\begin{align}\label{eq:MIshannon}
    \text{MI}(G_1,G_2)\approx {N\choose 2}[\text{H}_b(p_1)+\text{H}_b(p_2)-\text{H}_s(\textbf{\text{P}}_{12})],    
\end{align}
where 
\begin{align}
    \textbf{\text{P}}_{12}=\{p_{12},p_1-p_{12},p_{2}-p_{12},1-p_1-p_2+p_{12}\}    
\end{align}
is the (normalized) contingency table between the labellings
corresponding to $G_1$ and $G_2$---encoding the fraction of all
${N\choose 2}$ possible edge slots that are true positives, false
negatives, false positives, and true negatives respectively---and
$\text{H}_s$ is the Shannon entropy
\begin{align}
    \text{H}_s(\bm{p}) = -\sum_{i} p_i \log p_i.    
\end{align}
Typically, mutual information measures are written in terms of bits
per symbol rather than total bits (the units of
Eq.~(\ref{eq:MIshannon})). Dividing out the prefactor of ${N\choose 2}$,
we get the more familiar form of the mutual information
\begin{align}\label{eq:Igraph}
    \text{I}(G_1;G_2) = \text{H}_b(p_1)+\text{H}_b(p_2)-\text{H}_s(\textbf{\text{P}}_{12}),    
\end{align}
which just corresponds to the Shannon mutual
information~\cite{cover2012elements} between the binary vectors
encoding the edge positions in $G_1$ and $G_2$. 

The graph mutual information of Eq.~(\ref{eq:Igraph}) takes the form of
the standard Shannon mutual information, and therefore is bounded in
the interval $0\leq \text{I}(G_1;G_2)\leq
\text{H}_b(p_1),\text{H}_b(p_2)$, allowing for the normalized
expression 
\begin{align}\label{eq:NMIgraph}
    \text{NMI}(G_1;G_2) = 2\times \frac{\text{I}(G_1;G_2)}{\text{H}_b(p_1)+\text{H}_b(p_2)}.   
\end{align}
Equation~(\ref{eq:NMIgraph}) maps the graph mutual information to the
interval $[0,1]$ to allow for easier interpretation across systems of
different sizes. (There are many other options for normalizing the
mutual information, which have their own benefits and
drawbacks~\cite{vinh2010information}.)

We note that, as with any mutual information measure,
Eq.~(\ref{eq:Igraph}) is invariant to label permutations in the binary
representation of the graphs:
$\text{I}(G_1;G_2)=\text{I}(G_1;\overbar{G_2})=\text{I}(\overbar{G_1};G_2)=\text{I}(\overbar{G_1};\overbar{G_2})$,
where $\overbar{G_i}=\mathcal{G}-G_i$ is the graph complement of
$G_i$. This follows intuitively, since specifying the positions of
edges is equivalent to specifying the positions of non-edges in terms
of information content. In practice, however, this symmetry will
rarely ever have an effect on results, since we are nearly always in
the sparse regime where $p_1,p_2<0.5$, so the mutual information is
monotonic in the overlap $p_{12}$ for fixed $p_1,p_2$. 

Computing the graph mutual information measure in
Eq.~(\ref{eq:Igraph}) is very fast in practice, only requiring us to
find the sizes of the edge sets $G_1$, $G_2$, and $G_1\cap G_2$. The
total complexity of these calculations is equal to the complexity of
constructing the sets themselves, so poses no additional computational
burden.

We can adapt the mutual information of Eq.~(\ref{eq:Igraph}) to
networks with directed and/or self-edges by simply changing the
constant ${N\choose 2}$ for the size of $\mathcal{G}$ from which we
must pick the edges in our graph. For directed graphs with self-edges,
directed graphs without self-edges, and undirected graphs with
self-edges, we can set this constant to be $N^2$, $N(N-1)$, and
${N\choose 2}+N$ respectively. One can also adapt our network mutual
information framework to multigraphs by using the formulation we
present in the ``Mesoscale network mutual information'' section and
putting each node in their own group in the input partition.

One can also derive from Eq.~(\ref{eq:Igraph}) a variation of
information measure~\cite{meilua2003comparing} between graphs, thus
\begin{align}\label{eq:VIgraph} \text{VI}(G_1;G_2) =
\text{H}_b(p_1)+\text{H}_b(p_2) - 2\times\text{I}(G_1;G_2).
\end{align} Equation~(\ref{eq:VIgraph}) has an advantage over the
graph mutual information for various tasks such as network
embedding~\cite{GDC10} due to its pseudometric property, but we will
not explore its applications here.   

\subsection{Degree-corrected network mutual information}
\label{sec:degreecorrected}

The measure presented in the ``Network mutual information'' section is
based on an encoding scheme in which the receiver's knowledge of the
overlap (``true positive'' count) $E_{12}$ between the graphs $G_1$
and $G_2$ constrains the number of possibilities for $G_2$ once $G_1$
is known. This allows for a reduction in the information required to
specify $G_2$ after $G_1$ is sent, giving a mutual information
measure. However, one can also consider modifying the encoding process
to exploit other shared structure between the networks prior to the
transmission of $G_2$, which results in a mutual information that
captures a different notion of structural similarity between the
graphs. 

One such modification is ``degree-correction'', analogous to the
degree correction of the stochastic block model (SBM) for community
detection~\cite{peixoto2019bayesian} in which the degree sequence is
specified ahead of time to provide additional compression of network
data through its modular structure. Here, instead of specifying the
global edge overlap $E_{12}$ between $G_1$ and $G_2$, we can specify
how much overlap there is between the individual neighborhoods
$\partial^{(1)}_i$ and $\partial^{(2)}_i$ of each node $i$ in $G_1$
and $G_2$, respectively. In order to do this, a two-step process is
required: firstly, we must list the degree sequence
$\bm{k}_i=\{k_i(1),...,k_i(N)\}$ of each graph~$i$. Secondly, we must
specify how the total edge overlap $E_{12}$ is distributed among the
$N$ node neighborhoods.  These two steps have information content that
scales like $O(E_1+E_2)$, which for sparse graphs can be neglected
when we normalize by ${N\choose 2}$ to get the bits per symbol value
of the mutual information. We therefore ignore this intermediate
information cost, which we will see leads to a nice clean expression
for the degree-corrected graph mutual information. 
 
Given knowledge of the degrees $\bm{k}_2$ and how the $E_{12}$
overlapping edges are distributed across the node neighborhoods
(i.e.~the rows of $G_2$'s adjacency matrix representation), we can
compute the conditional entropy between $G_1$ and $G_2$ as
\begin{align}\label{eq:degcorrH21}
    \text{H}_{\text{deg}}(G_2\vert G_1) = \sum_{i=1}^{N}\log {k_1(i) \choose  k_{12}(i)}{N-k_1(i)\choose k_2(i)-k_{12}(i)},    
\end{align}
where $k_{12}(i)=\abs{\partial^{(1)}_i\cap \partial^{(2)}_i}$ is the
number of true positives (e.g.~overlapping edges with $G_1$) attached
to node~$i$. The first binomial coefficient in the summand counts the
number of possible configurations for the overlapping  edges
(i.e.~those shared with $G_1$) attached to node~$i$ in $G_2$.
Meanwhile, the second term counts the number of possible
configurations for the non-overlapping edges (i.e.~those not shared
with $G_1$) attached to node~$i$ in $G_2$. 

The analogous entropy expression for $G_2$ if the receiver does not
have knowledge of any of the shared structure with $G_1$---but does
know the degrees $\bm{k}_2$, which are independent of $G_1$---is given
by
\begin{align}\label{eq:degcorrH2}
    \text{H}_{\text{deg}}(G_2) = \sum_{i=1}^{N}\log {N \choose  k_2(i)}.   
\end{align}
This is just the amount of information required to specify the nodes
attached to $i$ given its degree. Technically both
Eqs.~(\ref{eq:degcorrH21}) and~(\ref{eq:degcorrH2}) are only upper bounds
on the conditional entropy and entropy for undirected graphs under
this encoding scheme, since only one edge direction must be known to
specify each edge. They are, however, exact for directed graphs in
which the degree sequences~$\bm{k}_i$ can be chosen to be either the
in- or out-degrees.

By comparing Eqs.~(\ref{eq:degcorrH2}) and~(\ref{eq:degcorrH21}) with
Eqs.~(\ref{eq:H2}) and~(\ref{eq:H21}), we can immediately identify the
mutual information value for this degree-corrected scheme as the
average of node-level mutual information values, thus
\begin{align}\label{eq:degcorrI}
\text{I}_{\text{deg}}(G_1;G_2) = \frac{1}{N}\sum_{i=1}^{N}[\text{H}_b(p_1(i))+\text{H}_b(p_2(i))-\text{H}_s(\textbf{\text{P}}_{12}(i))],    
\end{align}
where $p_1(i)=k_1(i)/(N-1)$, $p_2(i)=k_2(i)/(N-1)$, and
\begin{align}
\textbf{\text{P}}_{12}(i)&=\{p_{12}(i),p_1(i)-p_{12}(i),p_2(i)-p_{12}(i),\nonumber\\
&~~~~~~~~~~1-p_1(i)-p_2(i)+p_{12}(i)\},    
\end{align}
with $p_{12}(i)=k_{12}(i)/(N-1)$. We normalize by $N-1$ for graphs
without self-edges since a node can connect to at most $N-1$ nodes
excluding itself. For networks with self-edges, one can change the
normalization to $N$ to account for the possibility of a node that is
completely connected to all other nodes including itself.

A degree-corrected NMI measure can be constructed for
Eq.~(\ref{eq:degcorrI}) by noting that
$\frac{1}{2}[\text{H}_b(p_1(i))+\text{H}_b(p_2(i))]$ is an upper bound
for each summand
$\text{H}_b(p_1(i))+\text{H}_b(p_2(i))-\text{H}_s(\textbf{\text{P}}_{12}(i))$
in Eq.~(\ref{eq:degcorrI}), each of which is a mutual information
measure at the node neighborhood level. We can therefore replace
the summand in Eq.~(\ref{eq:degcorrI}) with this upper bound to
obtain an upper bound on the total degree-corrected mutual
information. The resulting normalized mutual information measure
is given by
\begin{align}\label{eq:DCNMI}
    \text{DC-NMI}(G_1;G_2) = 2\times \frac{\text{I}_{\text{deg}}(G_1;G_2)}{\frac{1}{N}\sum_i[\text{H}_b(p_1(i))+\text{H}_b(p_2(i))]},
\end{align}
which will be bounded to $[0,1]$. As before, we can also construct a
degree-corrected variation of information measure for networks, thus
\begin{align}
\text{VI}_{\text{deg}}(G_1;G_2) &= \frac{1}{N}\sum_{i=1}^{N}[\text{H}_b(p_1(i))+\text{H}_b(p_2(i))]\nonumber\\
&~~~~~~~~~-2\times \text{I}_{\text{deg}}(G_1;G_2). 
\end{align}
One can show that the above expression is also a pseudometric like the
usual variation of information. 

Similar to the non-degree-corrected mutual information measure of
Eq.~(\ref{eq:NMIgraph}), the degree-corrected mutual information
measures presented here can be computed with a time complexity that is
linear in the size of the edge sets $G_1$ and $G_2$ being compared.
Once the set intersection $G_1\cap G_2$ has been computed, we just
need to iterate through these shared edges again and increment
$k_{12}(i)$ whenever node~$i$ is present in the edge.

In contrast with the measures of the ``Network mutual information''
section, which are unaffected by where edges differ between the two
graphs, the degree-corrected measures presented here will ascribe
higher similarity to graphs whose edge discrepancies are concentrated
on relatively few nodes. In this sense, the DC-NMI of
Eq.~(\ref{eq:DCNMI}) focuses on node-level structural similarity,
while the NMI of Eq.~(\ref{eq:NMIgraph}) focuses on edge-level
similarity. We will see from experiments in Sec.~\ref{sec:results}
that this distinction becomes important for graphs with
heterogeneous degrees, since differences in network structure can
naturally become concentrated on high degree nodes, particularly
in the case of random edge rewiring. 

\subsection{Mesoscale network mutual information}
\label{sec:meso}

While the measures derived in the ``Network mutual information'' and
``Degree-corrected network mutual information'' sections will capture
the small-scale information shared between two graphs $G_1$ and $G_2$
(e.g.~the overlap of specific edges and node neighborhoods), they will
fail to capture mesoscale structure such as communities or
core-periphery structure unless the exact positions of the edges
within these larger-scale structures happen to overlap. Indeed, if
$G_1$ and $G_2$ are both sparse networks generated from an
SBM~\cite{holland1983stochastic}, there is a very low probability that
they share a substantial fraction of edges despite having very similar
mesoscale divisions into communities since each community is itself a
sparse random graph whose edge positions are uncorrelated. In this
case, the measures we have discussed will likely ascribe little
similarity to the two graphs despite their identical mesoscale
community structure. To obtain meaningful graph similarity results at
different scales of interest---particularly at the mesoscale where
community structure is present---it is therefore important to consider
network mutual information formulations that take larger-scale
structure into account while ignoring small-scale details. Here we
present one possible formulation of such a mutual information measure
between graphs which we call the \emph{mesoscale graph mutual
information}.

Consider the same problem setting as in the ``Network mutual
information'' section, where we have undirected, unweighted,
node-aligned graphs $G_1$ and $G_2$ with $N$ nodes and $E_1,E_2$
edges, respectively. Assume again that $N,E_1,E_2$ are already known
by the receiver. This time we will always allow $G_1$ and $G_2$ to
potentially have self- or multi-edges, since this will allow for
easier computations, but in principle the calculation can be extended
to graphs without self- or multi-edges with more detailed
combinatorics. If $G_1$ and $G_2$ share mesoscale structure (e.g.~in
the form of groups of highly connected nodes), but not necessarily
microscale structure (in the form of overlapping edges), then the
encoding schemes of the previous sections will be very inefficient,
since there is little shared information at the microscale we can
exploit to reduce the entropy of our transmission. However, if we
instead only aim for a mesoscale description of each network---such as
the edge counts within and between communities in a given partition of
the graphs---then we can formulate a mutual information measure that
captures the shared mesoscale information between $G_1$ and $G_2$
while ignoring their microscale differences.

Here we consider partitioning the nodes in both graphs $G_1$ and $G_2$
with the same (non-overlapping) partition $\bm{b}$ with $B$ groups. We
will denote with \mbox{$n_r=\sum_{i=1}^{N}\delta_{b_i,r}$} the number
of nodes with community label $r$ in the partition $\bm{b}$. The
partition $\bm{b}$ can be thought of as a specific coarse-graining of
the networks, and tuning $B$ allows us to interpolate between the
microscale ($B\sim \text{O}(N)$) and the macroscale ($B\sim
\text{O}(1)$) to capture similarities at the scale of interest. A
reasonable choice for $\bm{b}$ once the scale $B$ is chosen is a
community partition of one of the networks into $B$ groups, since this
represents a meaningful coarse-graining of the network at this scale.
However, in principle any choice for $\bm{b}$ is possible, and we will
show in Sec.~\ref{sec:results} that often one can choose meaningful
coarse-grainings $\bm{b}$ based on node metadata relevant to a
particular application. We will let $\bm{b}$ be known to the
receiver---its corresponding entropy term would vanish in our final
mutual information expression anyway. 

Under the node partition $\bm{b}$, each network $G_i$ can be described
by a coarse-grained representation $\tilde G_i^{(\bm{b})}$, defined as
the multiset of $E_i$ elements where element~$(r,s)$ with $r\leq s$
has a multiplicity~$m_i(r,s)$ equal to the number of edges in $G_i$
containing one node in group~$r$ and one node in group~$s$. (This is
equivalent to the mixing matrix representation of community-community
ties used in the microcanonical SBM~\cite{peixoto2012entropy}.)
Defining the scale of the full network $G_i$ to be order
$\text{O}(1)$, the representation $\tilde G_i^{(\bm{b})}$ captures the
aggregate structure of $G_i$ at a scale of order $\text{O}(B^{-1})$.
In the extreme case with $B=N$ (the partition $\bm{b}$ puts each node
into its own group), we have $\tilde G_i^{(\bm{b})}=G_i$ and we are
capturing network similarity at the scale $\text{O}(N^{-1})$. The
measure we present in the case $B=N$ can thus be used as a mutual
information between multigraphs $G_1$ and $G_2$.   

The mesoscale mutual information measure $\text{I}_{\bm{b}}(G_1;G_2)$
that we will derive aims to capture the amount of shared information
at the scale $\text{O}(B^{-1})$ between the graphs $G_1$ and $G_2$ by
computing the mutual information between the multisets $\tilde
G_1^{(\bm{b})}$ and $\tilde G_2^{(\bm{b})}$ for a chosen partition
$\bm{b}$. (As discussed, $\bm{b}$ can be derived by using network
structure itself or by using external metadata.) Instead of
formulating the mutual information from the perspective of conditional
entropy, we will motivate the mesoscale mutual information from a
\emph{joint} transmission process in which we transmit both $\tilde
G_1^{(\bm{b})}$ and $\tilde G_2^{(\bm{b})}$---first individually, then
together using their shared information. Formulations of mutual
information measures using conditional and joint entropies are
equivalent, but in our case the latter allows for a more
straightforward exposition. 

We can first consider transmitting each individual multiset, $\tilde
G_i^{(\bm{b})}$, separately to a receiver. Using the same calculation
procedure as with the entropy measures above, we can find these
individual entropies to be
\begin{align}\label{eq:Htil2}
    \text{H}(\tilde G_i^{(\bm{b})}) = \log \multiset{{B\choose 2}+B}{E_i},     
\end{align}
where 
\begin{align}\label{eq:multiset}
    \multiset{n}{k} = {n+k-1\choose k}    
\end{align}
is the multiset coefficient. In Eq.~(\ref{eq:Htil2}),
$\multiset{{B\choose 2}+B}{E_i}$ is the number of multisets $\tilde
G_i^{(\bm{b})}$ with $E_i$ elements one can construct from a set of
objects with cardinality ${B\choose 2}+B$, that is, the number of
independent combinations $(r,s)$ with $1\leq r\leq s\leq B$. The
logarithm of this quantity is thus the entropy of our encoding for
specifying $\tilde G_i^{(\bm{b})}$ given the constraints known by the
receiver, and transmitting the two multisets individually therefore
requires $\text{H}(\tilde G_1^{(\bm{b})})+\text{H}(\tilde
G_2^{(\bm{b})})$ bits of information. 

An important property of the multiset coefficient that we will use
when constructing our measure is that it is subadditive when
transformed by a logarithm to get an entropy. In other words, for any
$k,l$ we have
\begin{align}\label{eq:subadditivity}
\log \multiset{n}{k} + \log \multiset{n}{l} \geq  \log \multiset{n}{k+l}.  
\end{align}
To prove this, define $X^{(n,k)}$ as the set of non-negative integer
vectors of length $n$ whose values sum to $k$. The multiset
coefficient counts the number of unique vectors in $X^{(n,k)}$, i.e.
$\abs{X^{(n,k)}}=\multiset{n}{k}$. We can construct a map
$f:X^{(n,k)}\times X^{(n,l)}\to X^{(n,k+l)}$ given by $f(x,y)=x+y$.
The map $f$ is surjective, since any $z\in X^{(n,k+l)}$ has at least
one pair $(x,y)\in X^{(n,k)}\times X^{(n,l)}$ such that $f(x,y)=z$.
Therefore, $\multiset{n}{k}\multiset{n}{l}=\abs{X^{(n,k)}\times
X^{(n,l)}}\geq \abs{X^{(n,k+l)}}=\multiset{n}{k+l}$, and taking the
logarithm of both sides gives the result in
Eq.~(\ref{eq:subadditivity}).

Now we can consider transmitting $\tilde G_1^{(\bm{b})}$ and $\tilde
G_2^{(\bm{b})}$ together, exploiting the shared information between
them to reduce the total entropy of the transmission. Using the
generalization of the set intersection to
multisets~\cite{hein2003discrete}, we can define the true positives
$E_{12}^{(\bm{b})}$ in this context as
\begin{align}\label{eq:E12b}
    E_{12}^{(\bm{b})} &= \abs{\tilde G_1^{(\bm{b})} \cap \tilde G_2^{(\bm{b})}}\nonumber \\
                      &=\sum_{r\leq s}\text{min}(m_1(r,s),m_2(r,s)),
\end{align}
where the index~$r$ iterates over $1,\dots,B$, and the index~$s$ iterates
over $r,\dots,B$. Equation~(\ref{eq:E12b}) tells us the total number of common
group pairs $(r,s)$ (allowing duplicates) between the multisets
$\tilde G_1^{(\bm{b})}$ and $\tilde G_2^{(\bm{b})}$. The information
required to transmit $E_{12}^{(\bm{b})}$ is of \mbox{$\text{O}(\log
(E_1+E_2))$} and can be ignored. With $E_{12}^{(\bm{b})}$ known, there
are $\log \multiset{{B\choose 2}+B}{E_{12}^{(\bm{b})}}$ ways to
distribute the overlapping pairs among $\tilde G_1^{(\bm{b})}$ and
$\tilde G_2^{(\bm{b})}$. Once the overlapping pairs are known, we must
specify $E_i-E_{12}$ remaining pairs for each of the multisets $\tilde
G_i^{(\bm{b})}$. Putting this all together gives a total joint
information of
\begin{align}\label{eq:Hjointupper}
    \text{H}'=\log \multiset{{B\choose 2}+B}{E_{12}^{(\bm{b})}}\multiset{{B\choose 2}+B}{E_1-E_{12}^{(\bm{b})}}\multiset{{B\choose 2}+B}{E_2-E_{12}^{(\bm{b})}}.    
\end{align}

In principle, with the joint entropy of Eq.~(\ref{eq:Hjointupper}) one
can construct a mesoscale mutual information measure of
$\text{H}(\tilde G_1^{(\bm{b})})+\text{H}(\tilde
G_2^{(\bm{b})})-\text{H}'$. While this has some desirable properties,
it is not strictly increasing with the overlap $E_{12}^{(\bm{b})}$ due
to low overlap values placing substantial constraints on the multisets
$\tilde G_1^{(\bm{b})}$ and $\tilde G_2^{(\bm{b})}$ in some cases,
providing an efficient reduction in the entropy and a high mutual
information. Intuitively, one would expect that a higher overlap
$E_{12}^{(\bm{b})}$ among the multisets $\tilde G_1^{(\bm{b})}$ and
$\tilde G_2^{(\bm{b})}$ should result in a higher similarity value,
and according to this criterion the mutual information
$\text{H}(\tilde G_1^{(\bm{b})})+\text{H}(\tilde
G_2^{(\bm{b})})-\text{H}'$ is unsuitable as a candidate for a
similarity measure. Although the graph mutual information of
Eq.~(\ref{eq:Igraph}) technically also has a similar limitation due to
invariance under graph complements, as discussed in the ``Network
mutual information'' section this is irrelevant in the sparse regime
where $E_i<{N\choose 2}/2$. 

Instead of Eq.~(\ref{eq:Hjointupper}), it turns out that a lower bound
on the joint entropy can be used to produce a useful mutual
information similarity measure that does exhibit monotonicity with the
overlap $E_{12}^{(\bm{b})}$ as desired. Using the subadditivity of the
log-multiset coefficients in Eq.~(\ref{eq:subadditivity}), we have the
bound
\begin{align}
    \text{H}' \geq \log  \multiset{{B\choose 2}+B}{E_1+E_2-E_{12}^{(\bm{b})}}.   
\end{align}   
Using this bound on the joint entropy
\begin{align}
    \text{H}(\tilde G_1^{(\bm{b})}, \tilde G_2^{(\bm{b})}) = \log \multiset{{B\choose 2}+B}{E_1+E_2-E_{12}^{(\bm{b})}},    
\end{align}
we can now compute the mesoscale mutual information of graphs $G_1$
and $G_2$ with respect to partition $\bm{b}$ as
\begin{align}\label{eq:MImeso}
\text{I}^{(\bm{b})}_{\text{meso}}(G_1;G_2) &=  \text{H}(\tilde G_1^{(\bm{b})})+\text{H}(\tilde G_2^{(\bm{b})})-\text{H}(\tilde G_1^{(\bm{b})}, \tilde G_2^{(\bm{b})}).
\end{align}

For fixed $E_1,E_2$, we have that for any \mbox{$E_{12}^{(\bm{b})}\in [1,\min(E_1,E_2)]$} 
the mesoscale mutual information of Eq.~(\ref{eq:MImeso}) satisfies
\begin{align}\label{eq:increasing}
    \text{I}^{(\bm{b})}_{\text{meso}}(E_{12}^{(\bm{b})})-\text{I}^{(\bm{b})}_{\text{meso}}(E_{12}^{(\bm{b})}-1) 
    = \log \frac{\multiset{{B\choose 2}+B}{E_1+E_2-E_{12}^{(\bm{b})}+1}}{\multiset{{B\choose 2}+B}{E_1+E_2-E_{12}^{(\bm{b})}}}
    \geq 0,  
\end{align}
since the multiset coefficient $\multiset{n}{k}$ is a strictly
increasing function of $k$. Here we considered the mutual
information as a function of only the overlap~$E_{12}^{(\bm{b})}$
since this is the only remaining free parameter when $E_1,E_2$ are
fixed. Equation~(\ref{eq:increasing}) implies that, when all else is
constant, a greater overlap $E_{12}^{(\bm{b})}$ among $\tilde
G_1^{(\bm{b})}$ and $\tilde G_2^{(\bm{b})}$ gives a greater mesoscale
mutual information $\text{I}^{(\bm{b})}_{\text{meso}}(G_1;G_2)$, which
is consistent with what we expect from a similarity measure. 

From the previous argument, we can also see that
\begin{align}\label{eq:above0}
\text{I}^{(\bm{b})}_{\text{meso}}&\geq \text{I}^{(\bm{b})}_{\text{meso}}(E_{12}^{(\bm{b})}=0) \nonumber\\
    &=\log \frac{\multiset{{B\choose 2}+B}{E_1}\multiset{{B\choose 2}+B}{E_2}}{\multiset{{B\choose 2}+B}{E_1+E_2}} \nonumber \\
&\geq 0 ,
\end{align}
or in other words, the mesoscale mutual information is bounded below
by $0$. This follows from the subadditivity of the logarithm of the
multiset coefficient.

We can also use the inequality of Eq.~(\ref{eq:increasing}) to establish
an upper bound for the mesoscale mutual information. Without loss of
generality, label the graphs $G_i$ so that $E_1\leq E_2$. We then have
that the maximum overlap is $E_{12}^{(\bm{b})}=E_1$, and so
\begin{align}\label{eq:belowHavg}
    \text{I}^{(\bm{b})}_{\text{meso}}&\leq \text{I}^{(\bm{b})}_{\text{meso}}(E_{12}^{(\bm{b})}=E_1) \nonumber\\
&=\log \frac{\multiset{{B\choose 2}+B}{E_1}\multiset{{B\choose 2}+B}{E_2}}{\multiset{{B\choose 2}+B}{E_1+E_2-E_1}} \nonumber \\
&=\text{H}(\tilde G_1^{(\bm{b})}) \nonumber \\
&\leq \frac{\text{H}(\tilde G_1^{(\bm{b})})+\text{H}(\tilde G_2^{(\bm{b})})}{2},
\end{align}
where the last inequality uses the fact that the multiset coefficient
$\multiset{n}{k}$ is an increasing function of $k$ and that $E_1\leq
E_2$. 

Using the bounds of Eqs.~(\ref{eq:above0}) and~(\ref{eq:belowHavg}), we
can construct a normalized mesoscale mutual information
$\text{NMI}^{(\bm{b})}_{\text{meso}}$ as follows
\begin{align}\label{eq:mesoNMI1}
    \text{NMI}^{(\bm{b})}_{\text{meso}}(G_1;G_2) = 2\times \frac{\text{I}^{(\bm{b})}_{\text{meso}}(G_1;G_2)}{\text{H}(\tilde G_1^{(\bm{b})})+\text{H}(\tilde G_2^{(\bm{b})})},   
\end{align}
which mirrors the form of the graph NMI of Eq.~(\ref{eq:NMIgraph}). In
practice, it is also useful to consider an alternative normalization
with a tighter lower bound, thus
\begin{align}\label{eq:mesoNMI2}
    \text{MesoNMI}^{(\bm{b})}(G_1;G_2) = \frac{\text{I}^{(\bm{b})}_{\text{meso}}(G_1;G_2)-\text{I}^{(\bm{b})}_{\text{meso}}(E_{12}^{(\bm{b})}=0)}{\frac{\text{H}(\tilde G_1^{(\bm{b})})+\text{H}(\tilde G_2^{(\bm{b})})}{2}-\text{I}^{(\bm{b})}_{\text{meso}}(E_{12}^{(\bm{b})}=0)},    
\end{align}
where 
\begin{align}
    \text{I}^{(\bm{b})}_{\text{meso}}(E_{12}^{(\bm{b})}=0)=\log \frac{\multiset{{B\choose 2}+B}{E_1}\multiset{{B\choose 2}+B}{E_2}}{\multiset{{B\choose 2}+B}{E_1+E_2}}.    
\end{align}
Both Eqs.~(\ref{eq:mesoNMI1}) and~(\ref{eq:mesoNMI2}) will fall in the
range $[0,1]$, with
\mbox{$\text{NMI}^{(\bm{b})}_{\text{meso}}(G_1;G_2)=1$}
if and only if \mbox{$G_1=G_2$}. In our experiments,
we will use the mesoscale NMI measure defined in
Eq.~(\ref{eq:mesoNMI2}).

\begin{figure*}[ht!]
\includegraphics[width=0.975\textwidth]{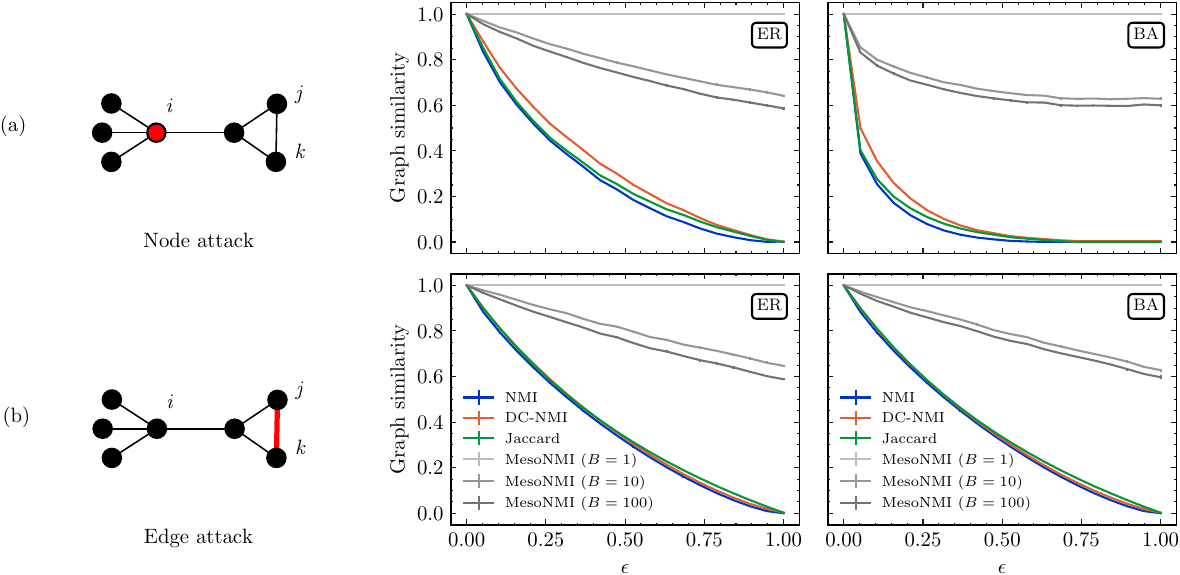}
\caption{
    \textbf{
    Graph similarity measures for networks under node and edge attacks.
    }
    (a)~Graph similarity as a function of the fraction of nodes
    attacked $\epsilon$ for random networks, where nodes are attacked
    in decreasing order of degree. Graph similarity is measured with
    the NMI, DC-NMI, and MesoNMI for $B\in \{1,10,100\}$ to capture
    multiple scales of similarity. The Jaccard index $\abs{G_1\cap
    G_2}/\abs{G_1\cup G_2}$ is included for comparison. The MesoNMI
    partitions $\bm{b}$ are computed with a standard stochastic block
    model (SBM) with fixed group sizes $B\in \{1,10,100\}$ on the
    initial (un-attacked) graph. Simulations are averaged over $10$
    realizations of the initial graph from the Erdős-Rényi model (ER,
    top left panel) and Barabási-Albert model (BA, top right panel)
    with $N=1000$ nodes and average degree $\langle k \rangle=10$
    (error bars indicate three standard errors and are vanishingly small). 
    (b)~Graph similarity as a function of the fraction of edges
    randomly rewired, for the same synthetic networks. Subtle
    differences in the decay rates of different similarity measures
    are reflective of intuitive properties of these measures, as
    discussed in Sec.~\ref{sec:results}.
}
\label{fig:fig1_er_ba}
\end{figure*}

The mesoscale mutual information above requires the user to choose a
common partition $\bm{b}$ of the networks $G_1$ and $G_2$, which
effectively sets the scale of interest for comparing the similarity of
the two graphs. One can choose the partition $\bm{b}$ in a number of
ways, a meaningful choice being a community partition of either $G_1$
or $G_2$ that decomposes the graph into densely connected groups of
nodes with sparser connections between groups. In principle, one can
also maximize or minimize $\text{I}^{(\bm{b})}_{\text{meso}}(G_1;G_2)$
over all possible partitions $\bm{b}$ with $B$ groups, which can
identify the divisions for which the networks are most/least similar
at the chosen scale.  Similarly, one could sample over partitions
$\bm{b}$ at a given scale $B$ and examine the full distribution of
values $\text{I}^{(\bm{b})}_{\text{meso}}(G_1;G_2)$ to get a
multifaceted assessment of the similarity between $G_1$ and $G_2$ at
the chosen scale. Finally, as we do in the experiments here, one can
use node metadata to define the partition $\bm{b}$ by grouping nodes
with similar characteristics.

All of the proposed measures highlight significant shared structure
because they are constructed to compute the amount of shared
information between two graphs $G_1$ and $G_2$ with respect to a
particular property of interest---individual edges for the NMI of
Eq.~\ref{eq:NMIgraph}, node neighborhoods for the DC-NMI of
Eq.~(\ref{eq:DCNMI}), and edge densities within groupings of nodes for
the MesoNMI of Eq.~(\ref{eq:mesoNMI2}). If there is redundant
information in the two graphs that can be compressed under an encoding
that exploits a particular property $X$ of interest, this indicates
that there is meaningful shared structure in the two graphs with
respect to $X$. The same property makes the proposed measures easily
interpretable: If one achieves a high NMI between two graphs, the
networks have a meaningful share of overlapping edges; if one achieves
a high DC-NMI between two graphs, the networks have a meaningful
amount overlap among individual node neighborhoods; and if one
achieves a high MesoNMI between two graphs, the networks have a
meaningful overlap in the ``super-edges'' of the coarse-grained
multigraph corresponding to the partition of interest. Finally, the
proposed methods are fast: The total runtime complexity of each of the
three proposed measures is $O(E_1+E_2)$, which is the same complexity
as creating the edge sets $G_1$ and $G_2$ in the first place.

Figure~\ref{fig:fig1_diagram} shows the application of all measures
presented in this section to a pair of example networks. 


\section{Results}
\label{sec:results}

In this section we apply our graph similarity measures in a variety of
experiments with synthetic and empirical networks. First, we
illustrate the differences between similarity measures by attacking
synthetic networks at different scales. We find that the different
encoding schemes indeed produce graph similarity measures that
highlight shared structure at different scales and are affected
differently by these structural perturbations. Then, we apply our
measures to a case study with an empirical multilayer network of
global trade patterns, finding that our measures capture meaningful
shared structure among the network layers representing the movement of
different goods.

\subsection{Similarity scores among perturbations of synthetic networks}
\label{sec:er-ba-sbm}

We examine the extent to which each of our similarity measures
captures different structural deviations by perturbing synthetic
reference networks with different noise (or ``attack'') strategies. In
each simulation, we first generate a reference network from a random
graph model---here we used random networks generated from the
Erdős-Rényi (ER) model~\cite{ER59,Gilbert59}, Barabási-Albert (BA)
model~\cite{BA99b}, and the stochastic block model (SBM) with equally
sized groups and mixing levels that only depend on whether the nodes
are in the same group or different
groups~\cite{holland1983stochastic,KN11a}. (This is also known as the
planted partition model.) These different reference graphs allow us to
examine the effects of global network structure and degree
heterogeneity on the similarity scores. We then perturb the reference
graph by attacking it with one of two types of moves: (1) node attacks
(Fig.~\ref{fig:fig1_er_ba}a), which take all the edges incident to a
node and rewire them uniformly at random to neighbors not currently
connected to the attacked node; and (2) edge attacks
(Fig.~\ref{fig:fig1_er_ba}b), which take an edge $(i,j)$ and place it
between a different pair of nodes $(i',j')$, chosen uniformly at
random from all nodes except $i,j$. We run our simulations by
attacking nodes in decreasing order of degree and edges in a random
order, measuring the extent to which the original network has been
perturbed with an attack fraction $\epsilon$ indicating the fraction
of nodes or edges that have been attacked.

It is worth noting that, for large $N$, the expected fraction of
shared edges $p_{12}$ is well approximated by $p_{12}\approx p_1p_2$,
and one can show that the resulting value of the NMI measure of
Eq.~(\ref{eq:NMIgraph}) is zero in this case. A similar argument can
be made for the DC-NMI, and this is seen numerically for $\epsilon=1$
in Figure \ref{fig:fig1_er_ba}(b).

Figures~\ref{fig:fig1_er_ba}a and~\ref{fig:fig1_er_ba}b show graph
similarity values for a number of measures versus the fraction
$\epsilon$ of node and edge attacks, respectively. We include the
Jaccard similarity $\abs{G_1\cap G_2}/\abs{G_1\cup G_2}$ for
reference. In these experiments, the partitions $\bm{b}$ required to
compute the MesoNMI measures from Eq.~(\ref{eq:mesoNMI2}) are computed
by fitting an SBM to the original (unperturbed) network with the
indicated number of groups $B$. Each curve is the average over 10
simulations, each starting from a different initial graph. All
networks in the experiment were of size $N=1000$ and had an average
degree of $\langle k \rangle=10$. Unless stated otherwise, error bars
indicate three standard errors.  The reference networks were generated
from an ER model (left panels) and a BA one (right panels), with the
goal of capturing the effects of degree heterogeneity on how each
similarity measure is penalized under node and edge attacks. 

In all panels, we see steeper decays in the similarity values of the
microscale measures---NMI, DC-NMI, and Jaccard index---than the
MesoNMI measures (except at $B=1$, which is trivially equal to $1$ for
all $\epsilon$ since the number of edges is constant throughout the
attack process). We can also see a greater (albeit still modest)
differentiation between the microscale measures for the node attacks
than the edge attacks. The standard NMI measure and Jaccard index are
unchanged between the two attack strategies, since it will be
penalized equally no matter how edges are replaced. However, the
\text{DC-NMI} is less penalized by the node attacks than the NMI or
the Jaccard index. This is because the rewiring of a hub node~$i$
primarily impacts the DC-NMI (Eq.~\ref{eq:DCNMI}) via the change in
this node's new neighborhood overlap $p_{12}(i)=0$, while $i$'s old
and new neighbors $j$ may see only slight changes to their values
$p_{12}(j)$ as these nodes often will have other neighbors that are
unchanged after the attack. In contrast, all the edges $(i,j)$
replaced due to the hub attack---which may constitute a substantial
fraction of all edges in the network---will penalize equally the total
edge overlap $\abs{G_1\cap G_2}$ considered by the NMI and Jaccard
measures.

We can also see from Fig.~\ref{fig:fig1_er_ba}a that degree
heterogeneity has a substantial effect on the sensitivity of the
similarity measures for the node attacks, since attacking nodes in
decreasing order of degree results in more edges being rewired at a
given attack fraction $\epsilon$ for heterogeneous (BA) than
homogeneous (ER) degree distributions. However, this effect is not
present in Fig.~\ref{fig:fig1_er_ba}b for the edge attacks due to all
edges providing a roughly equal contribution to each similarity
measure.

The MesoNMI measures in all panels present fairly uniform patterns,
with the decay in similarity becoming more severe as we increase the
number of groups $B$ in the partition $\bm{b}$ of the reference
network. This is consistent with the attacks being performed at the
microscale (e.g. on nodes and edges) rather than at the mesoscale:
under the random rewiring of individual edges, the mixing structure
within and between groups of the node partition remains less affected
since these perturbations may produce new edges that run between the
same pairs of groups. This effect is more pronounced as we decrease
the number of groups $B$, since with few groups it is more likely that
rewiring an edge $(i,j)$ to a new edge $(i',j')$ will result in the
same pair of groups $b_i,b_j$ appearing on the ends of the edge. This
robustness to microscale attacks also manifests in less sensitivity to
the degree heterogeneity of the original graph, as can be seen in the
top right panel.

\begin{figure}[h!]
\includegraphics[width=0.450\textwidth]{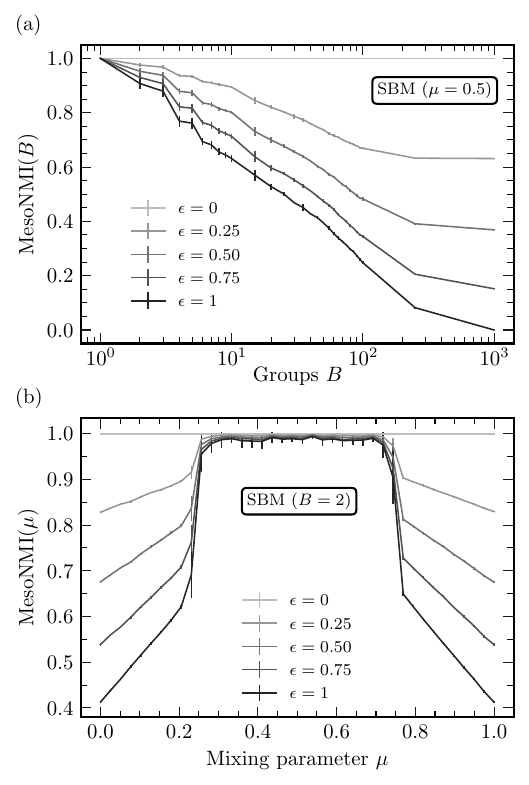}
\caption{
    \textbf{
    Mesoscale mutual information between stochastic block model (SBM) networks.
    }
    Edge attack simulations were performed on networks generated from
    an SBM with two groups of $500$ nodes, average degree of $10$, and
    mixing level $\mu \in [0,1]$ fixing the fraction of edges running
    between nodes of the same group identity. 
    (a)~MesoNMI values for different edge attack fractions $\epsilon$
    (indicated by curves of shades of gray) as a function of the
    number of groups $B$ of the node partition $\bm{b}$ used for the
    MesoNMI calculation, for SBM networks with no mixing preference
    ($\mu=0.5$, equivalent to Erdős-Rényi random graphs). 
    The MesoNMI is more sensitive to edge-level attacks as the scale
    of interest for comparison gets smaller (i.e. the number of groups
    $B$ gets larger).
    (b)~MesoNMI as a function of the mixing level $\mu$ of the initial
    graph being attacked, with the partition $\bm{b}$ used for the
    MesoNMI calculations being fixed as the initial graph's planted
    partition into $B=2$ groups. As the mixing level moves away from
    $\mu=0.5$, we see a stronger dependence of the MesoNMI on the
    attack level $\epsilon$ due to the rewiring of inter-community
    ($\mu=0$) or intra-community ($\mu=1$) edges to produce an
    equitable mixture of these two edge types in expectation. 
}
\label{fig:fig2_sbm}
\end{figure}

Figure~\ref{fig:fig2_sbm} shows an edge attack experiment, except we
use an initial graph drawn from an SBM with two equal groups and
tunable mixing level $\mu$ fixing the fraction of edges that run
within groups. The value of $\mu=0.5$ corresponds to no mixing
preference between the two groups (i.e. the network is an ER random
graph), while $\mu=0$ corresponds to completely disassortative
structure in which all edges are between groups, and $\mu=1$
corresponds to a completely assortative structure in which all edges
are among nodes that have the same group affiliation. As before, we
fix the number of nodes to be $N=1000$ and average degree of $\langle
k\rangle=10$.  For the MesoNMI measures, we calculate the partitions
$\bm{b}$ in the same way as in Fig.~\ref{fig:fig1_er_ba}, computing
the SBM-optimal partition on the reference network with the desired
value of $B$ being fixed. Although the reference SBMs actually only
have two groups in their planted structure, setting $B$ to different
values for the attack experiments allows us to examine the similarity
of the perturbed networks with this reference graph at different
scales using the MesoNMI measure. 

\begin{figure*}[t!]
\includegraphics[width=1.0\textwidth]{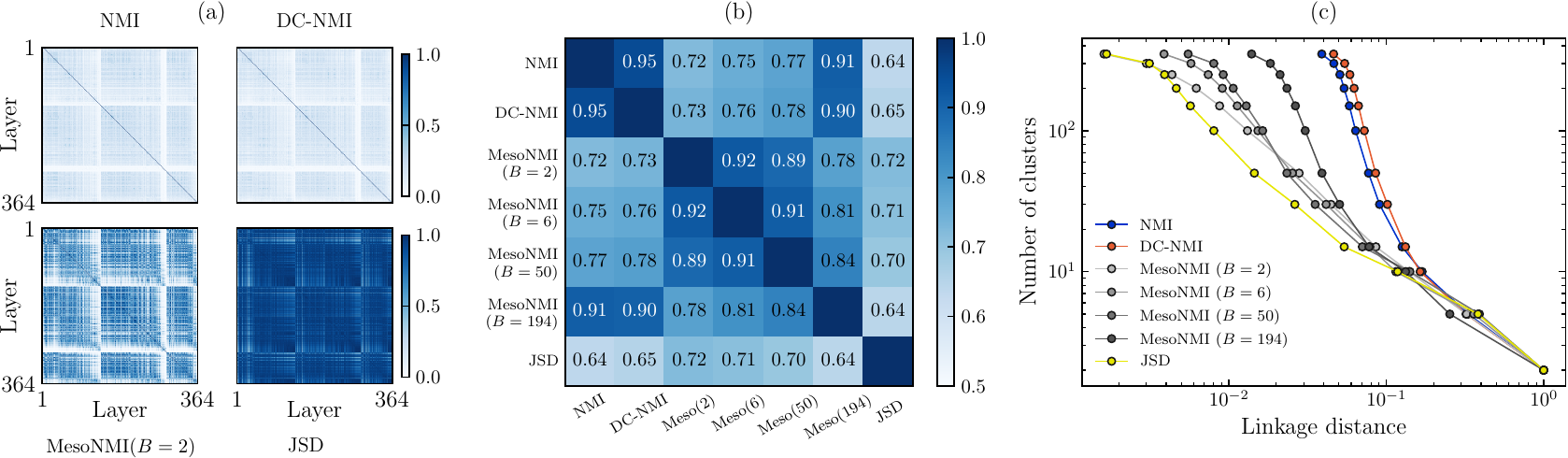}
\caption{
    \textbf{
    Comparison of graph similarity values among layers of the FAO trade network. 
    }
    (a)~Pairwise similarity matrices among layers of the FAO trade
    network~\cite{de2015structural}, each layer representing the
    global trade patterns among countries for a particular good. The
    MesoNMI was computed with respect to a partition of the country
    nodes in each layer according to a Global North-South
    dichotomy~\cite{globalsouth}.  The network Jensen Shannon
    divergence (JSD) measure of~\cite{de2015structural} is transformed
    into a similarity measure using $1-\text{JSD}$ and included for
    comparison. All matrices indicate a similar block structure to the
    layer similarities, and as the network scale of interest increases
    (NMI to DC-NMI, to MesoNMI to JSD) we find systematically higher
    similarity values, with the MesoNMI having the greatest
    discriminative power. 
    (b)~Rank-biased overlap (RBO)~\cite{webber2010similarity} between
    the pairwise distances calculated using each pair of similarity
    measures. For example, the (NMI, DC-NMI) entry of this matrix is
    the RBO between the entires of the top two panels in (a). As the
    scale of interest decreases, we find greater RBO between the
    corresponding pairwise distance matrices. 
    (c)~Number of clusters versus the corresponding Ward linkage
    distance for a hierarchical clustering of the
    layers~\cite{de2015structural}.  There are discrepancies in the
    hierarchical cluster structure inferred using the measures, with
    measures operating at similar scales having similar linkage
    patterns.
}
\label{fig:fig4_fao}
\end{figure*}

In Fig.~\ref{fig:fig2_sbm}a we plot the MesoNMI versus the number of
groups $B$ for different attack fractions $\epsilon$ (shades of gray)
for an SBM with no mixing structure ($\mu=0.5$, i.e.~the reference
network is an ER random graph).  Simulations for the first
interval $B \in [1, 10]$ were averaged over $100$ random initial
networks, while for the tail end we used $10$ simulations in
total.  We find (as expected) that the MesoNMI decreases
monotonically as a function of $\epsilon$ for all values of $B$,
with $\epsilon=0$ trivially providing no change to the similarity
scores. We also find, as in Fig.~\ref{fig:fig1_er_ba}, that for a
fixed value of the attack fraction $\epsilon$ the MesoNMI values
decrease monotonically as a function of $B$, indicating that the
microscale edge attacks are not felt as intensely by the measures
that aggregate structural information at larger scales. 

In Fig.~\ref{fig:fig2_sbm}b we plot the MesoNMI versus the mixing
parameter $\mu$ used to generate the reference SBM graph, also at
different attack fractions $\epsilon$. In this case, the partition
$\bm{b}$ used to compute the MesoNMI is the original planted partition
of the reference network into two groups (simulations were averaged
over 50 random initial networks).  We find that, for a given attack
fraction $\epsilon$, the sensitivity in the MesoNMI is strongly
dependent on the mixing level $\mu$, with values decreasing as we move
away from $\mu=0.5$. This is because edge attacks will affect a highly
(dis)assortative partition more than one with little mixing
preference, since edges are more likely to be rewired to new group
affiliations if they are highly concentrated among nodes with certain
pairs of group affiliations. We also observe an interesting phase
transition-like behavior in the MesoNMI at roughly $\mu=0.25$ and
$\mu=0.75$, which may have qualitative similarities with the
detectability transition of the SBM~\cite{decelle2011asymptotic} in
which community structure is suddenly statistically indistinguishable
from random edge placement at a certain level of mixing.

These experiments altogether verify our intuition about how each
measure values certain deviations in network structure, as well as
confirm that our mesoscale mutual information measure is truly
capturing variation at coarser scales than the NMI and DC-NMI
measures. We supplement these tests with further experiments using
other random graph ensembles in Supplementary Figure~\ref{fig:figS1_lim}.  

\subsection{FAO trade network case study}
\label{sec:fao}

To examine how the proposed measures can be used to extract meaningful
summaries of shared structure in sets of networks, we apply our
measures to the multilayer FAO network of global trade
patterns~\cite{de2015structural,tantardini2019comparing}, where nodes
represent $N=214$ countries, an edge $(i,j)$ represents the trade of a
good between countries $i$ and $j$, and each layer (of which there are
$364$) represents a different good being traded.  We kept only the
countries with gross domestic product (GDP) accessible through the
World Bank indicators~\cite{worldbank2010gdp} of 2010 (the same year
the tradings took place), and so the final network studied was reduced
to $194$ nodes.  The network was binarized and converted to an
undirected representation, facilitating a straightforward comparison
with the network Jensen Shannon divergence (JSD) measure proposed
in~\cite{de2015structural} and evaluated on the same dataset.
(However, as discussed in the ``Mesoscale network mutual information''
section, one can apply all of our measures to the original directed
network representation, and apply the MesoNMI measures to the weighted
representation by considering it as a multigraph.) The network JSD
measure of~\cite{de2015structural} is computed using the spectrum of
the combinatorial Laplacian, and should capture both small- and
large-scale structural similarities between networks as with other
spectral measures~\cite{apolloni2011introduction,wilson2008study}.

We applied the NMI (Eq.~\ref{eq:NMIgraph}), DC-NMI
(Eq.~\ref{eq:DCNMI}), MesoNMI (Eq.~\ref{eq:mesoNMI2}), and the network
JSD~\cite{de2015structural}---which was converted to a similarity
measure $1-\text{JSD}$ to facilitate a direct comparison with our
measures---to compare all pairs of layers in the FAO trade network and
perform hierarchical clustering analysis as
in~\cite{de2015structural}. For the MesoNMI, we used a number of
partitions $\bm{b}$ reflecting meaningful divisions of the countries
at multiple scales: (1)~a~partition into $B=2$ groups representing the
Global North and Global South in accordance to the United Nations'
Finance Center for South-South Cooperation~\cite{globalsouth};
(2)~a~partition into $B=6$ groups representing the continents in the
dataset; (3) a partition into $B=50$ equally sized groups of countries
after ordering by GDP---this is to facilitate an analysis at an
intermediate network scale; and (4) a partition of the network into
$B=194$ groups each of size one---this is to compare and contrast the
multigraph-based encoding of the MesoNMI with the graph-based encoding
of the standard NMI, which can give substantially different results in
practice. A diagram of our preprocessing and analysis for this section
is shown in Fig.~\ref{fig:fig-fao-diagram}.

In Fig.~\ref{fig:fig4_fao} we show the results of these experiments.
In particular, panel~\ref{fig:fig4_fao}a shows the pairwise
similarity matrices for a few of the measures, which convey a similar
block structure for the network layer comparisons. However, as the
network scale of interest increases (NMI to DC-NMI, to MesoNMI to JSD)
we can observe increasing similarity values, reflecting greater
overall similarity among the networks when viewing them at larger
scales. The MesoNMI measures find a much greater variability in the
similarity values across network layers, as indicated by the range of
shades present in the heatmap, while the NMI, DC-NMI, and JSD measures
infer a tighter range of similarity scores among the layers. 

We then compare these similarity matrices to examine how similar the
ranks of the entries are across pairs of measures. We use the
rank-biased overlap (RBO)~\cite{webber2010similarity}, a measure
indicating the overlap in the rankings of values from two lists. The
RBO provides an alternative measure to the Spearman rank correlation
that more heavily weights the contributions from the highest ranked
items in both lists, which in our case allows us to more heavily favor
pairs of measures that assign high similarity scores to common pairs
of layers. In Fig.~\ref{fig:fig4_fao}b we plot the results of applying
the RBO to the flattened similarity matrices of each pair of measures
studied. We observe a very clear trend in which the RBO of a pair of
measures is inversely related to the difference in their scale of
interest. For example, the NMI and DC-NMI results have a decreasing
RBO with the MesoNMI results as we move from $B=194$ to $B=50$, to
$B=6$ to $B=2$, and the similarity among the MesoNMI results exhibits
the same trend. Meanwhile, the highest RBO between any of our measures
and the JSD is $0.72$ (for MesoNMI($B=2$)), which is in turn equal to
the lowest of the RBO values among any of our measures (NMI and
MesoNMI($B=2$)). This may reflect the fact that the JSD is capturing
both small- and large-scale similarities among the networks, while our
measures are targeting particular scales of interest. In
Fig.~\ref{fig:figS_spearmann_ami}a we repeat the experiment in
Fig.~\ref{fig:fig4_fao}b using Spearman rank correlation instead of
RBO, finding the same qualitative results. 

We then examine the behavior of all measures in a hierarchical
clustering context. We consider the rows of the similarity matrices in
Fig.~\ref{fig:fig4_fao}a as node embeddings that reflect the positions
of each layer relative to the other layers. We then cluster the layers
hierarchically using the Ward linkage criterion, as done
in~\cite{de2015structural} for the JSD distance measure. In this
formulation, the pair of clusters whose layers have the least
discrepancy in their similarity patterns with other clusters will be
iteratively merged until there is only one remaining cluster of
layers---the full multilayer network. We show the linkage results from
this experiment in Fig.~\ref{fig:fig4_fao}c. We can see that, as in
Fig.~\ref{fig:fig4_fao}b, the measures are ordered roughly by their
scale of interest, with measures operating at similar scales having
similar linkage patterns. Measures operating at small scales (e.g.~NMI
and DC-NMI) have a relatively high linkage threshold at which they
begin to merge clusters, while measures operating at coarser scales
(e.g. JSD and MesoNMI($B=2$)) have a relatively low linkage threshold
at which they begin to merge clusters. At large numbers of clusters,
this pattern is reflective of the increasing similarity scores seen
across scales in Fig.~\ref{fig:fig4_fao}a. However, we see many of the
linkage patterns for the coarser measures cross over those for the
small-scale measures at roughly $10$ clusters. This indicates that
these measures will give qualitatively distinct clustering patterns
beyond those resulting simply from a systematic difference in the
magnitude of their similarity scores.

To further examine the discrepancies in the clustering structure
induced by each similarity measure, we test the extent to which the
clusters obtained through hierarchical clustering reflect similarities
in the products being traded in each layer. We assign a product type
to each layer using the following $12$ categories: \{Proteins, Grains,
Dairy, Fruits, Vegetables, Sweets, Drinks, Spices, Animals, Raw
Materials, Tobacco, Other\}~\cite{kirkley2023compressing}. We call
these the ``ground truth'' product types for simplicity, but emphasize
that, since we are performing unsupervised learning, we are not trying
to exactly capture the distinctions among product types with our
method (a task for which supervised techniques are better suited). 

\begin{table}[t!]
    \centering
    \setlength{\arrayrulewidth}{0.75pt}
    \begin{tabular}{|@{\hspace{1.5mm}} r @{\hspace{2.5mm}} | @{\hspace{2mm}} c @{\hspace{2mm}} | @{\hspace{2mm}} c @{\hspace{2mm}} | @{\hspace{3mm}} l @{\hspace{2mm}}|} 
        \hline
        $\,$ & $\dfrac{\langle\text{Sim}^{(\text{betw})}\rangle}{\langle\text{Sim}^{(\text{with})}\rangle}$ & $\dfrac{\sigma(\text{Sim}^{\text{(betw)}})}{\sigma(\text{Sim}^{\text{(with)}})}$  & AMI \\
        \hline    
        NMI              & 0.810 & 0.842 &  0.175 \\
        DC-NMI           & 0.805 & 0.817 &  0.223 \\
        MesoNMI($B=2$)   & 0.935 & 1.007 &  0.031 \\ 
        MesoNMI($B=6$)   & 0.929 & 0.998 &  0.055 \\
        MesoNMI($B=50$)  & 0.922 & 0.994 &  0.041 \\
        MesoNMI($B=194$) & 0.864 & 0.924 &  0.138 \\
        JSD              & 0.982 & 1.100 &  0.040 \\
        \hline
    \end{tabular}
    \caption{
        \textbf{
        Correspondence between similarity scores and product types of FAO trade network layers.
        }
        The layers of the FAO trade network were each assigned to one
        of 12 ``ground truth'' product categories following the
        classification in~\cite{kirkley2023compressing}. The clusters
        for each method were compared with the ground truth categories
        using the ratio of the average between-category similarity
        and the average within-category similarity, with lower
        values indicating more tightly knit categories according
        to the similarity method of interest. The relative
        standard deviations of these similarity scores are also
        included for additional context. Then, the similarity
        matrices computed in Fig.~\ref{fig:fig4_fao} were
        clustered using Ward linkage as
        in~\cite{de2015structural}, with the number of clusters
        fixed at 12 to ensure a fair comparison across methods.
        The adjusted mutual information
        (AMI)~\cite{vinh2010information} between the ground truth
        layer partition and the inferred layer partition was
        computed for each method.  At face value, both tests
        indicate that the micro-scale measures are capturing more
        of the similarity among layers in the same category than
        the measures focusing on the meso- and macro-scales,
        suggesting that scattered individual edges may carry more
        information about shared global trade patterns than
        larger-scale network structure.
    }
    \label{tab:f_ami_scores}
\end{table}

We compare the clustering outputs from each measure with these ground
truth categories using two different measures. First, we examine the
extent to which each measure finds greater similarity among product
layers within the same category versus other categories. To do this we
compute the ratio of the average similarity scores among layers within
the same ground truth category
($\langle\text{Sim}^{(\text{with})}\rangle$) and layers within
different ground truth categories
($\langle\text{Sim}^{(\text{betw})}\rangle$). A lower ratio indicates
that the measure is better discriminating among the ground truth
categories, since it finds higher similarities among within-category
layers than between-category layers. The results for this experiment
are shown in the left column of Table~\ref{tab:f_ami_scores}. We can
see that the NMI and \text{DC-NMI} tend to best discriminate among the
product categories, while the JSD is the poorest discriminator among
the categories, with similarity scores that are nearly equal between
and within categories. We also compute the adjusted mutual information
(AMI)~\cite{vinh2010information} between the partitions of the layers
induced by each similarity measure with the ground truth layer
partition according to product category. Since there are many
strategies for determining where to cut each measure's corresponding
clustering dendrogram to find its associated partition of the layers,
we cut each measure's dendrogram at $12$ clusters---the same as the
number of clusters in the ground truth partition---to facilitate a
fair comparison across measures. We plot the results in the right
column of Table~\ref{tab:f_ami_scores}. We find that the NMI and
DC-NMI find clusters that are the most similar to the ground truth
according to the AMI, consistent with their capability to distinguish
these categories in the previous experiment. Similarly, we find poor
performance for the JSD and MesoNMI measures on this dataset.  At face
value, these results indicate that the similarity among product
networks within the same category manifests at the microscale rather
than the meso- or macro-scales. However, these results may vary
depending on preprocessing choices for the networks (e.g. network
pruning), which can affect the similarity values.  In
Fig.~\ref{fig:figS_spearmann_ami}b we compute the AMI between the
inferred clusters and the ground truth categories for all levels of
each measure's hierarchical linkage dendrogram, finding roughly the
same ordering of the measures as for the cut at $12$ clusters. In
Fig.~\ref{fig:figS_within_between} we also plot the distribution of
similarity scores within and between ground truth product groups to
visualize the full distributions of scores rather than just the
averages reported in Table~\ref{tab:f_ami_scores}. Finally, in
Fig.~\ref{fig:fig_clusters} we plot the distribution of ground truth
categories within the inferred layer clusters for a few of the
measures discussed for a more detailed picture. 

In Figs.~\ref{fig:figS3_aps}~and~\ref{fig:figS4_airport} we apply our
measures to two additional multilayer network datasets representing
collaboration patterns among scientists within different fields of
physics and routes for different airline companies
respectively~\cite{nicosia2015measuring}. 


\section{Conclusion}

In this paper we have proposed a family of mutual information measures
for computing the similarity between a pair of node-aligned networks.
We adapt the encodings used to construct these measures to
accommodate structural similarity at multiple scales within the
network. By applying the proposed measures in a range of tasks, we
find that the proposed measures are able to consistently capture
meaningful notions of network similarity at the desired scale
under perturbations to synthetic networks, as well as capture
heterogeneity among the layers in real multilayer network datasets
arising in the study of global trade, scientific collaboration,
and transportation.

There are a number of ways in which our measures can be extended in
future work. Firstly, we can enable analyses in a broader range of
contexts by lifting the restriction of the method to node-aligned
graphs (at the potential cost of increased computational burden),
refining our encoding for weighted graphs to enable meaningful
analyses beyond the multigraph representation, and/or extending our
pairwise encodings to populations of networks (e.g. as in
\cite{kirkley2023compressing}). The MesoNMI measure we propose does
not consider the impact of degree heterogeneity on similarity, and so
could also in principle be extended to a degree-corrected version as
done with the standard graph NMI measure. One can additionally apply
new encodings under our framework to capture similarity with respect
to the presence of complex local structures such as motifs or other
subgraphs while allowing for network of different sizes, similar to
the calculations performed in~\cite{coupette2021graph}. Lastly, by
adapting the combinatorial calculations to a new set of constraints
the methods here presented can in principle be extended to more
general discrete structures, such as hypergraphs or simplicial
complexes, to account for higher-order
interactions~\cite{battiston2020networks}. There are also many avenues
of exploration to examine additional applications of our methods for
downstream tasks such as network regression or anomaly detection and
to a wider range of network datasets such as connectomics
data~\cite{coupette2022differentially}.


\section*{Acknowledgments}
\vspace{-\baselineskip}
A.K. was supported by an HKU Urban Systems Fellowship Grant and the
Hong Kong Research Grants Council under Grant no. ECS--27302523.  
F.B. acknowledges support from the Air Force Office of Scientific
Research under award number FA8655-22-1-7025.


\section{Data and code Availability}

Data and code implementing the measures presented in this work are available at
\url{https://github.com/hfelippe/network-MI} 
with the DOI~\href{https://zenodo.org/doi/10.5281/zenodo.13145601}{\texttt{10.5281/zenodo.13145602}}.

%
%
%

\section{Author Contributions}

H.F.: conceptualization, software, validation, formal analysis,
investigation, data curation, writing—original draft, visualization; 
F.B.: conceptualization, writing—review and editing, supervision,
funding acquisition; 
A.K.: conceptualization, methodology, software, writing—original
draft, writing—review and editing, supervision.
All authors gave final approval for publication and agreed to be held
accountable for the work performed therein.


\bigskip

\section{Competing Interests}

The authors declare no competing interests.


%
%

\clearpage

\onecolumngrid

\begin{center}
  \textbf{
      \large Supplementary Information 
}\\[.2cm]
\end{center}

\setcounter{equation}{0}
\setcounter{figure}{0}
\setcounter{table}{0}
\setcounter{page}{1}
\setcounter{section}{0}
\renewcommand{\theequation}{S\arabic{equation}}
\renewcommand{\thefigure}{\textbf{S\arabic{figure}}}
\renewcommand{\thetable}{S\arabic{table}}
\renewcommand{\thepage}{S\arabic{page}}
\renewcommand{\thesection}{Note S\arabic{section}}
\renewcommand{\bibnumfmt}[1]{[#1]}
\renewcommand{\citenumfont}[1]{#1}

\section{Within-ensemble graph similarity}

\begin{figure*}[h!]
\centering
\includegraphics[width=1.0\textwidth]{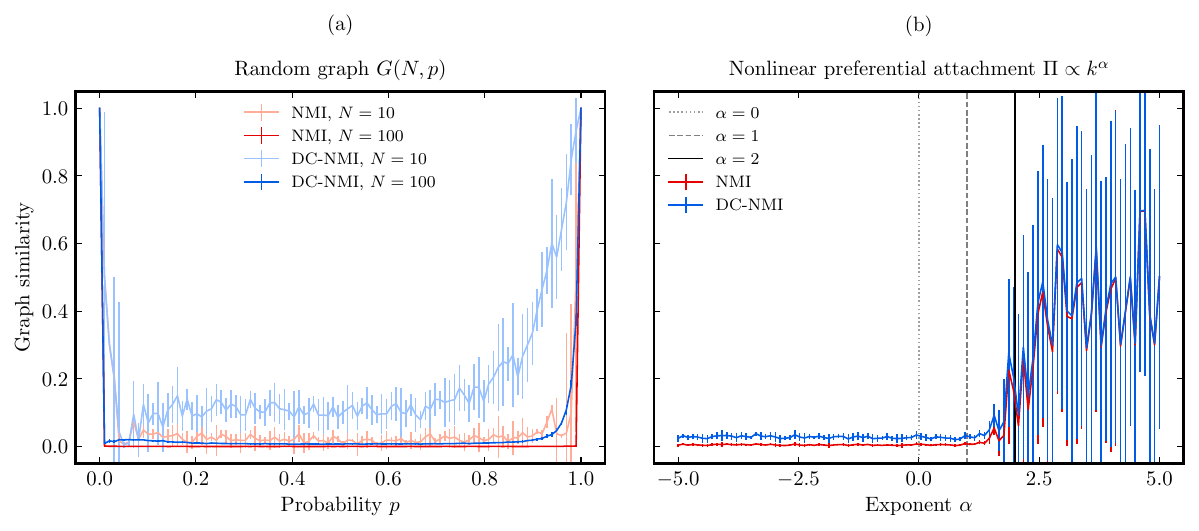}
\caption{
    \textbf{Average similarity among graphs generated from random graph ensembles.}
    (a) Average pairwise graph similarity for $100$ samples from the
    Erdos-Renyi random graph model with $N=10,100$ nodes and variable
    connection probability $p$. Error bars indicate 
    one standard error in the mean. 
    (b) Repeating the experiment but sampling networks from a
    nonlinear preferential attachment network growth model with
    exponent $\alpha$, we find almost no similarity between sampled
    graphs for $\alpha<2$, followed by a sudden transition at
    $\alpha=2$ to highly fluctuating similarity values. This is due to
    the predominance of loopy graphs during the $\alpha < 2$ regime,
    which results in low values of edge overlap between graphs, while
    for $\alpha>2$ we have star-like networks rooted at each of the
    two seed nodes with equal probability, resulting in a
    similarity value that is either $\approx 0$ or $\approx 1$ for
    any given pair of networks.
}
\label{fig:figS1_lim}
\end{figure*}

\clearpage

\section{Robustness checks for FAO trade network experiments}

\begin{figure*}[ht!]
\centering
\includegraphics[width=\textwidth]{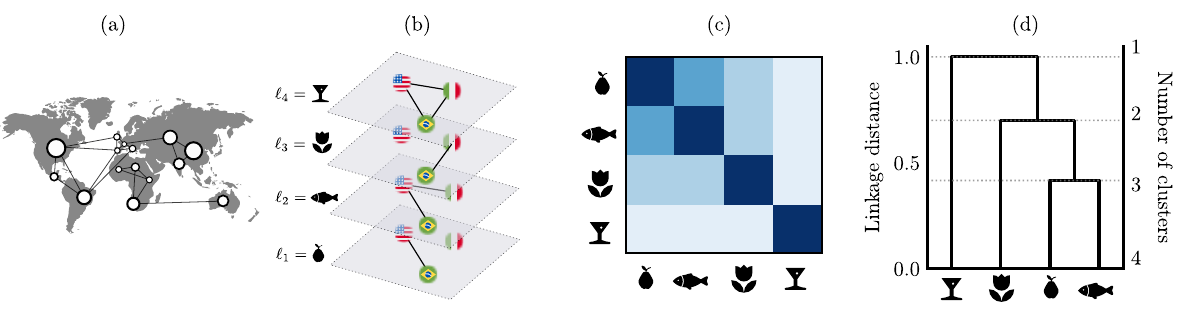}
\caption{
    \textbf{Preprocessing and analysis of the multilayer FAO trade network.}
    (a)~The FAO trade network consists of $N=194$ countries (nodes)
    exchanging a total of $L=364$ products (layers).  Two countries
    are connected within a layer $\ell$ if they traded any volume of
    product $\ell$.
    (b)~We illustrate the network's multilayer structure using four
    product layers $\ell$ depicting fruits, fish, plants, and
    beverages.  For simplicity, we only show trades among three
    countries: Brazil, Italy, and the United States.  Due to different
    trading patterns, there is great variability in the number of
    edges for each product layer, while the number of nodes is fixed
    in order to satisfy the node-aligned condition of our measures.
    (c)~Pairwise similarity matrices of dimension $L\times L$ are
    computed among all layers for each of the six graph similarity
    measures considered. 
    (d)~Hierarchical clustering is applied over each similarity matrix
    by iteratively merging the clusters of products with the greatest
    average similarity. The linkage distance allows us to assess how
    the six measures differ from each other (see Fig.~\ref{fig:fig4_fao}).
}
\label{fig:fig-fao-diagram}
\end{figure*}

\begin{figure}[h!]
\includegraphics[width=0.790\textwidth]{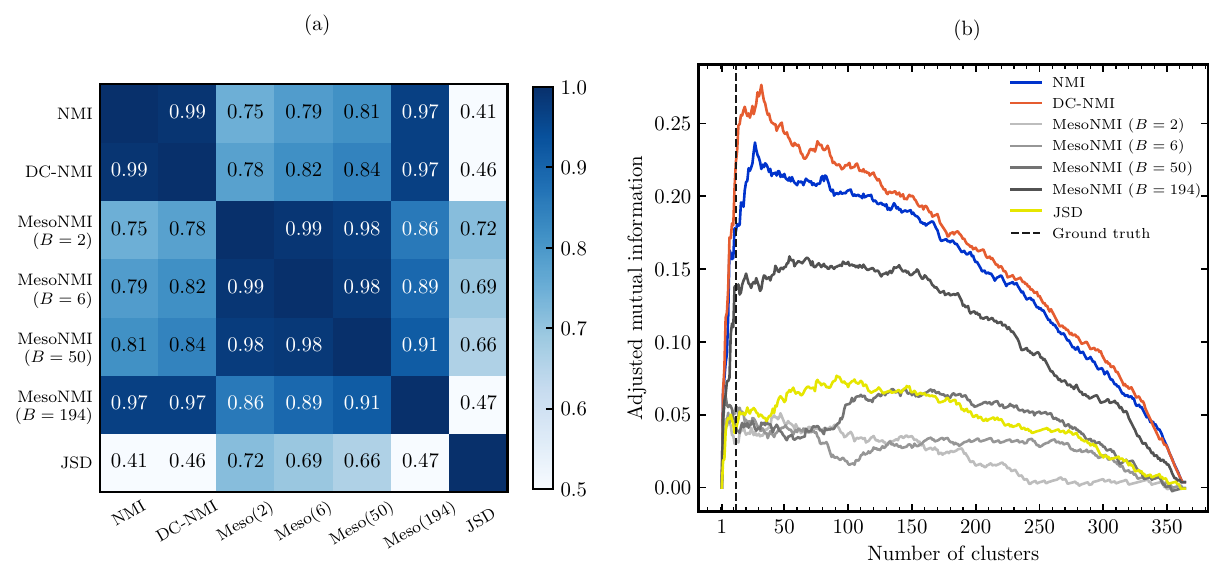}
\caption{
    \textbf{
    Spearman correlation and AMI for FAO layer similarity scores.
    }
    (a)~The same test as in Fig.~\ref{fig:fig4_fao}b was repeated but
    using Spearman rank correlation instead of rank-biased overlap
    (RBO).  The same qualitative results can be observed.
    (b)~The AMI scores in Table~\ref{tab:f_ami_scores} were computed
    using all levels of each measures cluster hierarchy. The dashed
    line indicates $12$ clusters, which is the level at which the
    measures are compared in Table~\ref{tab:f_ami_scores}. We can
    observe that all measures more or less maintain the same ordering
    regardless of the level at which we choose to cut the dendrogram. 
}
\label{fig:figS_spearmann_ami}
\end{figure}

\begin{figure}[hb!]
\includegraphics[width=0.800\textwidth]{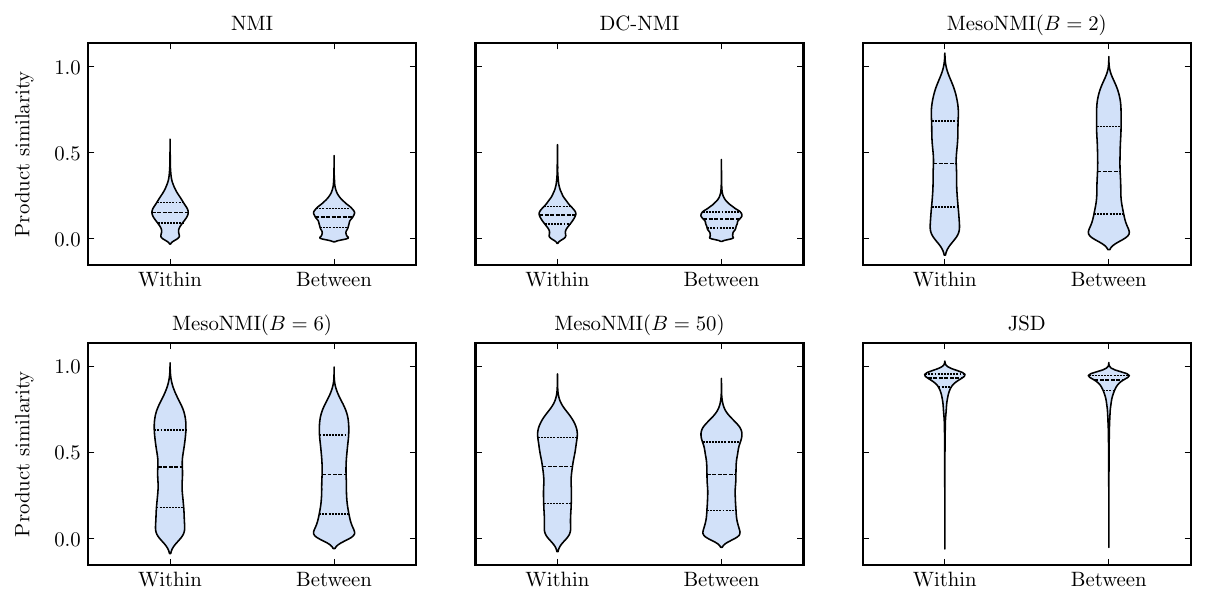}
\caption{
    \textbf{
    Distributions of similarity scores among product layers of the FAO trade network, within and between ground truth product categories.
    }
    Violin plots of the similarity scores for the FAO trade network
    layers used to compute
    $\langle\text{Sim}^{(\text{between})}\rangle/\langle\text{Sim}^{(\text{within})}\rangle$
    in Table~\ref{tab:f_ami_scores}. Dashed lines indicate the
    quartiles of the score distribution. As seen in
    Fig.~\ref{fig:fig4_fao}a, we observe the greatest variance in the
    scores of the MesoNMI measures, and as observed in
    Table~\ref{tab:f_ami_scores} we see the greatest relative
    discrepancy in the within-cluster and between-cluster
    distributions for the NMI and DCNMI measures. 
}
\label{fig:figS_within_between}
\end{figure}

\begin{figure*}[h!]
\centering
\includegraphics[width=\textwidth]{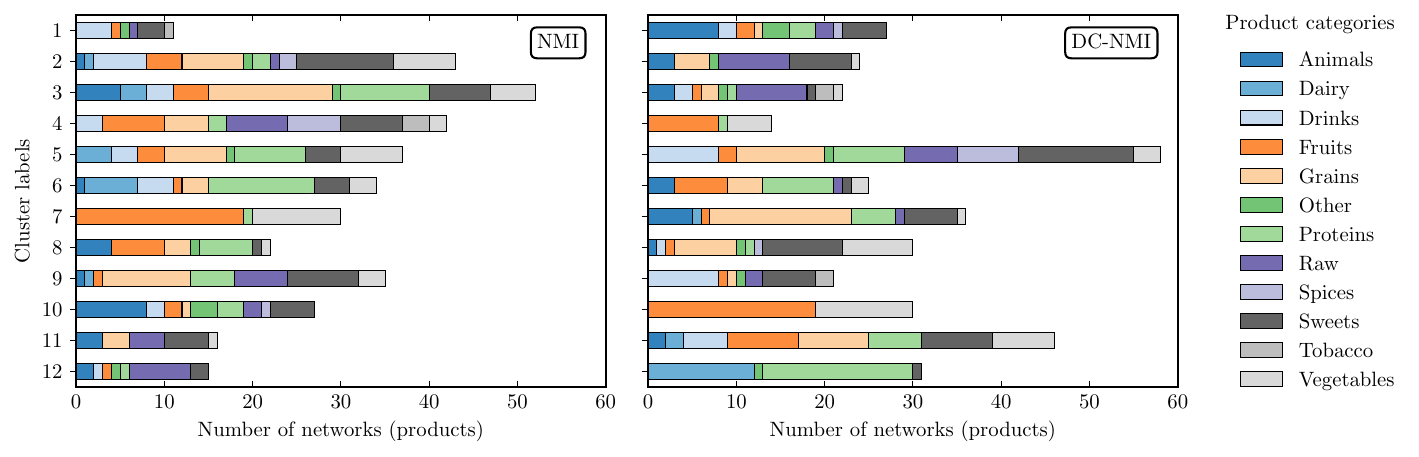}
\caption{
    \textbf{Distribution of ground truth categories for each inferred cluster.}
    Clusters inferred for the FAO network with each method are
    compared against ground truth product categories, allowing us to
    compare the true and inferred labels as well as variations in the
    cluster composition among the methods. 
}
\label{fig:fig_clusters}
\end{figure*}

Here we run a number of additional tests to confirm the findings 
in Sec.~\ref{sec:results} for the FAO trade network, which
is illustrated in Fig.~\ref{fig:fig-fao-diagram} together with a
diagram of our preprocessing and analysis.
Figure~\ref{fig:figS_spearmann_ami} repeats key experiments
in Sec.~\ref{sec:results} with different evaluation strategies, and
Fig.~\ref{fig:figS_within_between} plots the distributions of layer
similarity scores for products in the same ground truth category and
products in different ground truth categories.  Finally,
Fig.~\ref{fig:fig_clusters} shows where the ground truth labels differ
from our inferred clusters by plotting the distribution of ground
truth products within each inferred cluster for multiple methods.

\clearpage

\section{Scientific collaboration and continental airport networks}

To further test our graph similarity measures on empirical data, we
applied them to both a network of scientific collaboration across
different fields of physics and a network of continental
airports~\cite{nicosia2015measuring}.

\begin{figure}[hb!]
\includegraphics[width=0.90\textwidth]{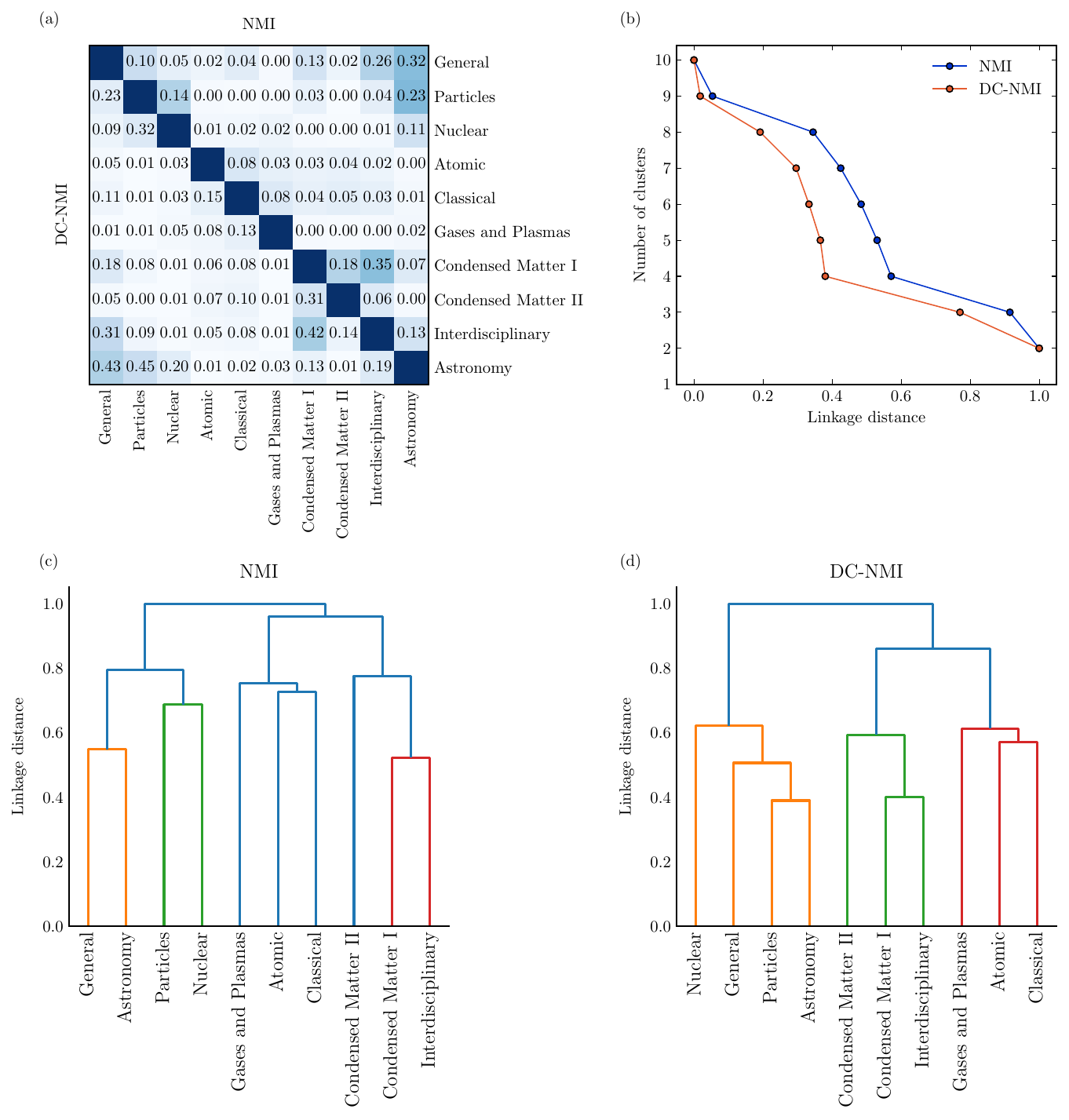}
\caption{
    \textbf{
    Similarity between fields of physics according to the NMI and DC-NMI measures.
    }
    (a)~Similarity scores between different fields of physics defined
    according to their PACS number. 
    (b)~Hierarchical clustering applied to the NMI and DC-NMI distance
    matrices in panel (a). The rapid decay in linkage distance of the
    DC-NMI measure indicates a higher similarity at the author (node)
    level between fields of physics (network layers). 
    (c)~NMI and (d)~\text{DC-NMI} dendrograms for hierarchical
    clustering with respect to similarity value. Both measures extract
    clusters that are consistent with topical overlap among the
    fields.  
}
\label{fig:figS3_aps}
\end{figure}

The American Physical Society (APS) scientific collaboration network
consists of authors (nodes) that have published at least one journal
paper together (edges) in any of the ten highest-level categories
(layers) in the Physics and Astronomy Classification Scheme~(PACS).
Figure~\ref{fig:figS3_aps} shows the NMI and DC-NMI scores among these
different PACS networks.

\clearpage

\begin{figure}[ht!]
\includegraphics[width=1.00\textwidth]{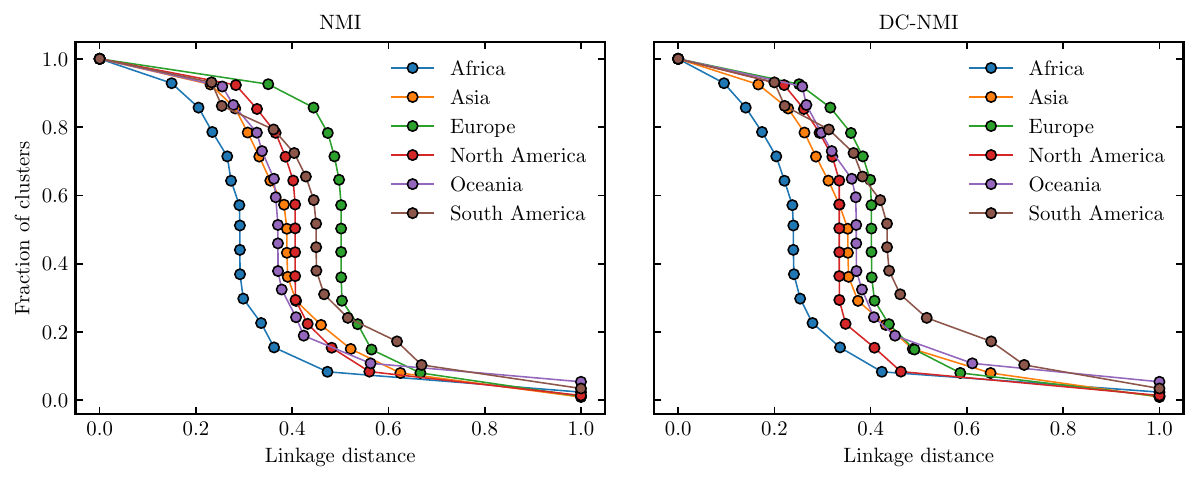}
\caption{
    \textbf{
    Similarity score cluster dendrograms among continental airports.
    }
    Hierarchical clustering was applied to the NMI and DC-NMI distance
    matrices obtained from the pairwise computation of similarity
    between airline networks within each continent. We can see that in
    both cases Africa has the greatest cluster differentiation at the
    small scale, while the ordering of the linkage patterns for the
    NMI and DC-NMI measures differ for the other continents.  
}
\label{fig:figS4_airport}
\end{figure}

The OpenFlights continental airport network consists of airports
(nodes) that have at least one flight between them (edges) operated by
the same airline company (layers). We test the similarity between
airlines from each of the six continents using the NMI and DC-NMI
measures (see Fig.~\ref{fig:figS4_airport}).

\end{document}